\documentclass{article}
\usepackage{amssymb}
\usepackage{bm}

\usepackage{graphicx}
\usepackage{amsmath}

\def\nsection#1{\section{#1}\setcounter{equation}{0}}

\newcommand{\bE}{\boldsymbol{E}}
\newcommand{\bR}{\boldsymbol{R}}
\newcommand{\qq}{\begin{eqnarray}}
\newcommand{\qqq}{\end{eqnarray}}
\newcommand{\ee}{{\rm e}}

\newcommand{\CF}{{\cal F}}

\newcommand{\CI}{{\cal I}}
\newcommand{\CJ}{{\cal J}}
\newcommand{\CO}{{\cal O}}

\newcommand{\CR}{{\cal R}}

\newcommand{\CW}{{\cal W}}

\newcommand{\sfx}{{\mathsf x}}
\newcommand{\sfy}{{\mathsf y}}
\newcommand{\sfX}{{\mathsf X}}
\newcommand{\sfW}{{\mathsf W}}

\newcommand{\sfQ}{{\mathsf Q}}
\newcommand{\sfU}{{\mathsf U}}

\newcommand{\hf}{\frac{_1}{^2}}
\newcommand{\antivec}[1]{\ensuremath\reflectbox{\hspace{-0pt}$\vec{\text{\reflectbox{$\!#1\!\,$}}}\,$}}

\begin{document}

\title{{\bf FLUCTUATION\ \,RELATIONS\ \,FOR\ \,DIFFUSION\ PROCESSES}}

\author{Rapha$\rm\ddot{e}$l Chetrite\ \,\ and \,
Krzysztof Gaw\c{e}dzki\footnote{Member of C.N.R.S.}\\ 
\\ Universit\'e de Lyon, C.N.R.S., ENS-Lyon, Laboratoire de Physique, 
\\ 46 All\'ee d'Italie, 69364 Lyon, France}


\maketitle

\abstract{\noindent The paper presents a unified approach 
to different fluctuation relations for classical nonequilibrium dynamics 
described by diffusion processes. Such relations compare the statistics 
of fluctuations of the entropy production or work in the original process 
to the similar statistics in the time-reversed process. The origin 
of a variety of fluctuation relations is traced to the use of 
different time reversals. It is also shown how the application 
of the presented approach to the tangent process describing the 
joint evolution of infinitesimally close trajectories of the original 
process leads to a multiplicative extension of the fluctuation 
relations.}
\vskip 2cm

\nsection{Introduction}

Nonequilibrium statistical mechanics attempts a statistical
description of closed and open systems evolving under the action of 
time-dependent conservative forces or under time-independent or time 
dependent non-conservative ones. {\bf Fluctuation relations} are robust 
identities concerning the statistics of entropy production or performed 
work in such systems. They hold arbitrarily far from thermal equilibrium. 
Close to equilibrium, they reduce to Green-Kubo or fluctuation-dissipation 
relations, usually obtained in the scope of linear response theory 
\cite{Zwan,HaySas}. Historically, the study of fluctuation relations 
originated in the numerical observation of Evans, Cohen and Morriss 
\cite{ECM} of a symmetry in the distribution of fluctuations of microscopic 
pressure in a thermostatted particle system driven by external shear. 
The symmetry related the probability of occurrence of positive and 
negative time averages of pressure over sufficiently long time intervals
and predicted that the former is exponentially suppressed with respect
to the latter. Ref.\,\cite{ECM} attempted to explain this observation 
by a symmetry,  induced by the time-reversibility, of the statistics 
of partial sums of finite-time Lyapunov exponents in dissipative dynamical 
systems. This was further elaborated in \cite{EvSear} where an argument 
was given explaining such a symmetry in a transient situation when one starts 
with a simple state which evolves under dynamics, see also \cite{EvSear2}. 
In refs.\,\cite{GC,GC1}, Gallavotti and Cohen provided a theoretical 
explanation of the symmetry observed numerically in \cite{ECM} employing 
the theory of uniformly hyperbolic dynamical systems. In this theory, 
the stationary states correspond to invariant measures of the SRB type 
\cite{LSY} and the entropy production is described by phase-space 
contraction \cite{Ruelle1}. The authors of \cite{GC,GC1} established 
a {\bf fluctuation theorem} about the rate function describing the 
statistics of large deviations of the phase-space contraction in the 
time-reversible dynamics. To relate to the behavior of realistic systems, 
they formulated the {\bf chaotic hypothesis} postulating that many such 
systems behave, for practical purposes, as the uniformly hyperbolic ones. 
They interpreted the numerical observations of ref.\,\cite{ECM} as a 
confirmation of this hypothesis. The difference between the fluctuation 
relations for a transient situation analyzed in \cite{EvSear,EvSear2} 
and the stationary one discussed in \cite{GC,GC1} was subsequently 
stressed in \cite{CG}. The debate about the connection between the 
transient and stationary fluctuation relations still continues, see 
e.g. \cite{SRE} and \cite{Gall3}. 
\vskip 0.1cm

In another early development, Jarzynski established in \cite{Jarz1} a simple 
transient relation for the statistics of fluctuations of work performed 
on a system driven by conservative time-dependent forces. This relation
is now known under the name of Jarzynski equality. A similar observation, 
but with more limited scope, was contained in the earlier work 
\cite{BK1,BK2,BK3}, see \cite{Jarz6} for a recent comparison. The simplicity 
of the Jarzynski equality and its possible applications to measurements 
of free-energy landscape for small systems attracted a lot of attention, 
see \cite{Ritort,Ritort1} and the references therein. 
\vskip 0.1cm

The first studies of fluctuation relations dealt with the deterministic 
dynamics of finitely-many degrees of freedom. Such dynamics may be also 
used to model systems interacting with environment or with heat reservoirs. 
To this end, one employs simplified finite-dimensional models of reservoirs 
forced to keep their energy constant \cite{EvMorr}. This type of models 
was often used in numerical simulations and in discussing fluctuation 
relations, see e.g.\,\cite{Gall3}. A more realistic treatment of reservoirs 
would describe them as infinite systems prepared in the thermal equilibrium 
state. Up to now, only infinite systems of non-interacting particles could 
be treated effectively, see \cite{EPRB1,EPRB2}. A less realistic description 
of interaction with environment or with reservoirs consists of replacing 
them by a random noise, usually shortly correlated in time. This leads 
to Markovian stochastic evolution equations. Stochastic models are often 
easier to control than deterministic ones and they became popular in modeling 
nonequilibrium dynamics.
\vskip 0.1cm

In \cite{Jarz2}, Jarzynski generalized his relation to time-dependent 
Markov processes with the instantaneous generators satisfying the detailed 
balance relation. At almost the same time, Kurchan has shown in 
\cite{Kurchan} that the stationary fluctuation relations hold for the 
stochastic Langevin-Kramers evolution. His result was extended to more 
general diffusion processes by Lebowitz and Spohn in \cite{LebowSp}. 
In \cite{Maes}, Maes has traced the origin of fluctuation relations 
to the Gibbsian nature of the statistics of the dynamical histories, see 
a recent discussion of fluctuation relation from this point of view 
in \cite{Maes1}. Searles and Evans generalized there transient fluctuation 
relation to the stochastic setup in \cite{SE}. Finally, within the 
stochastic approach, the scope of the transient fluctuation relations 
was further extended due to the works of Crooks \cite{Crooks1,Crooks2}, 
Jarzynski \cite{Jarz4}, Hatano and Sasa \cite{HH}, Speck and Seifert 
\cite{SpS} and Chernyak, Chertkov and Jarzynski \cite{ChChJarz}, just to 
cite only the papers that influenced most the present authors. It is worth 
stressing that the general transient fluctuation relations do not impose 
the time reversibility of the dynamics but compare the fluctuation 
statistics of the original process and of its time reversal. Such 
an extension of the scope of fluctuation relations is a possibility 
in the stationary case as well, but it becomes a necessity in many 
transient situations. Within the theory of the hyperbolic dynamical 
systems, the stationary fluctuation theorem of \cite{GC} was recently 
generalized to the random dynamics in \cite{BonGal}.  
\vskip 0.1cm

In \cite{BFF}, Balkovsky Falkovich and Fouxon noticed another
robust relation concerning the large deviations of finite-time 
Lyapunov exponents in the context of homogeneous hydrodynamic flows.
It was remarked in \cite{FGV}, that this observation, which we
shall call, following \cite{GawWarw}, the {\bf multiplicative fluctuation 
relation}, provides an extension of the previously known fluctuation 
relations for the phase-space contraction. The simple argument presented 
in \cite{BFF} dealt with a transient situation. It was very similar to 
the original Evans-Searles argument as formulated later in \cite{EvSear2}. 
The multiplicative fluctuation relation was explicitly checked in 
the Kraichnan model of hydrodynamic flows \cite{BFF,FGV,aniso}.  
\vskip 0.1cm

The theoretical work on fluctuation relations has established most
of them as mathematical identities holding within precisely defined 
models, but concerning statistics of events that are rare, especially 
for macroscopic systems. The relevance of such identities to numerical 
simulations and, even more, to real experiments, required a confirmation. 
Numerical (see e.g. \cite{BGallG,HJarz,vZonC,GZGall,WEvSear}) and 
experimental testing of the fluctuation relations (see e.g. 
\cite{CilibL,GCilib,CRJSTB,BandiCG,JGDPC,IPPRS}) has attracted over
years a lot of attention, inspiring further developments. It will 
probably remain an active field in the future. It is not, however, 
the topic of the present paper.  
\vskip 0.1cm

The growing number of different fluctuation relations made urgent a 
development of a unifying approach. Several recent reviews partially 
provided such a unification from different points of view, see 
ref.\,\cite{EvSear2,Maes1,Kurchan1,ChChJarz}. In the present paper, we 
attempt another synthesis, with the aim of supplying a uniform derivation 
of most of the known fluctuation relations, including the multiplicative 
ones. We shall work in the setup of (possibly non-autonomous) diffusion 
processes in finite-dimensional spaces, somewhat similar, but more general 
that the one adopted in \cite{LebowSp}. The systems considered include, 
as special cases, the deterministic dynamics, the Langevin stochastic 
equation, and the Kraichnan model of hydrodynamic flow. This is certainly 
not the most general setup possible for discussing fluctuation relations 
(for example, the discrete-time dynamics, the stochastic dynamics with 
jumps, or non-Markovian evolutions are not covered), but it is general 
enough for a unified discussion of a variety of aspects of fluctuation 
relations. Most of our considerations are simple extensions of 
arguments that appeared earlier in usually more constrained contexts.
There are two basic ideas that we try to exploit to obtain a larger 
flexibility than in the previous discussions of fluctuation relations. 
The first one concerns the possible time-reversed processes that we admit. 
This idea appeared already in \cite{ChChJarz}, where two different time 
inversions were used for the Langevin dynamics with non-conservative forces, 
leading to two different backward processes and two different fluctuation 
relations. We try to exploit the freedom of choice of the time-inversion 
in a more systematic way. The second idea, which seems original to us, 
although it is similar in spirit to the first one, is to obtain new 
fluctuation relations by considering new diffusion processes derived from 
the original one. In particular, we show that the multiplicative fluctuation 
relations for general diffusion processes may be obtained by writing a more 
standard relation for the tangent diffusion process describing a simultaneous 
evolution of infinitesimally close trajectories of the original process. 
The same idea may be used \cite{ChHorGaw} to explain additional fluctuation 
relations, like the one for the rate function of the difference of 
finite-time Lyapunov exponents ``along unstable flag'' that was 
observed in \cite{aniso} for the anisotropic Kraichnan model.  
\vskip 0.2cm

The present paper is organized as follows. In Sect.\,\ref{sec:forw},
we define the class of diffusion processes that will be discussed
and list four special cases. Sect.\,\ref{sec:trprob} recalls the
notions of transition probabilities and generators of a diffusion
process, as well as the detailed balance relation. In 
Sect.\,\ref{sec:tanproc}, we introduce the tangent diffusion process
induced form the original one and define the phase-space contraction.
Time inversions leading to different backward processes 
are discussed in Sect.\,\ref{sec:back}, with few important examples 
listed in Sect.\,\ref{sec:extimer}. A formal relation between 
the expectations in the forward and in the backward process is
introduced in Sect.\,\ref{sec:forwback}. As examples, we discuss
the case of tangent process in the homogeneous Kraichnan flow,
a simple generalization of the detailed balance relation and the $1^{st}$ 
law of thermodynamics for the Langevin dynamics. Sect.\,\ref{sec:Jarz} 
is devoted to a general version of the Jarzynski equality, whose different 
special cases are reviewed, and Sect.\,\ref{sec:SpSeif} to a related 
equality established by Speck and Seifert in \cite{SpS}. We formulate 
the Jarzynski equality as a statement that for a certain functional 
$\,\CW\,$ of the diffusion process, the expectation value of 
$\,\ee^{-\CW}\,$ is normalized. In Sect.\,\ref{sec:entrprod}, the functional 
$\,\CW\,$ is related to the entropy production and the positivity of its 
expectation value is interpreted as the $2^{\rm nd}$ law of thermodynamics 
for the diffusive processes. In Sect.\,\ref{sec:LR}, we show how the general 
Jarzynski equality reduces in the linear response regime to the Green-Kubo 
and Onsager relations for the transport coefficient and to the 
fluctuation-dissipation theorem. In Sect.\,\ref{sec:1D}, we discuss 
briefly a peculiar one-dimensional Langevin process in which the 
equilibrium is spontaneously broken and replaced by a state with a 
constant flux, leading to a modification of the fluctuation-dissipation 
relation. The model is well known from the theory of one-dimensional 
Anderson localization and describes also the separation of infinitesimally 
close particles with inertia carried by a one-dimensional Kraichnan flow. 
Sect.\,\ref{DFR} formulates in the general setup of diffusion processes 
what is sometimes termed a detailed fluctuation relation 
\cite{Jarz4,Crooks2}, an extension of the Crooks fluctuation relations
\cite{Crooks1}. Few special cases are retraced in Sect.\,\ref{sec:spcases}. 
\vskip 0.1cm

Up to this point of the paper, the discussion is centered on the 
transient evolution where the system is initially prepared 
in a state that changes under the dynamics. In Sect.\,\ref{sec:stationFR}, 
we discuss the relation of the transient fluctuation relations to the 
stationary ones which pertain to the situation where the initial state 
is preserved by the evolution. The stationary relations are usually
written for the rate function of large deviations of entropy production
observed in the long-time regime. In our case, they describe the long
time asymptotics of the statistics of $\,\CW$. \,The Gallavotti-Cohen 
relation was the first example of such identities. We show how the 
fluctuation relation for the tangent process in the homogeneous 
Kraichnan flow discussed in Sect.\,\ref{sec:forwback} leads
to a generalization of the Gallavotti-Cohen relation that involves
the large-deviations rate function of the so called stretching
exponents whose sum describes the phase-space contraction. In 
Sect.\,\ref{sec:MFR}, we extend such a multiplicative fluctuation 
relation to the case of general diffusion processes. Sect.\,\ref{sec:Npt} 
contains speculation about possible versions of fluctuation relations 
for multi-point motions and Sect.\,\ref{sec:concl} collects our
conclusions. Few simple but more technical arguments are deferred
to Appendices in order not to overburden the main text, admittingly
already much more technical than most of the work on the subject.
Some of the technicalities are due to a rather careful treatment 
of the intricacies related to the conventions for the stochastic 
differential equations that are usually omitted in the physical 
literature. The aim at generality, even without pretension of
mathematical rigor, places the stress on the formal aspects and 
makes this exposition rather distant from physical discourse, 
although we make an effort to include many examples that illustrate 
general relations in more specific situations. The physical content 
is, however, more transparent in examples to such examples which 
are scarce in the present text but which abound in the existing 
literature to which we often refer. Certainly, the paper will be too 
formal for many tastes, and we take precautions to warn the potential 
reader who can safely omit the more technical passages.  
\vskip 0.2cm

\noindent{\bf Acknowledgements}. \,The authors are grateful to
S. Ciliberto, G. Falkovich, I. Fouxon, G. Gallavotti and P. Horvai 
for discussions. 
\vskip 0.3cm

\nsection{Forward process}
\label{sec:forw}

As mentioned in Introduction, the present paper deals with
non-equilibrium systems modeled by diffusion processes
of a rather general type. More concretely, the main objects of 
our study are the stochastic processes $\,\sfx_t\,$ in $\bR^{d}$ 
(or, more generally, on a $d$-dimensional manifold), described 
by the differential equation
\qq
\dot{x}\ =\ u_t(x)\,+\,v_t(x)\,,  
\label{E0}
\qqq
where $\,\dot{x}\equiv\frac{dx}{dt}$ \,and,  on the right hand side, 
$\,u_t(x)\,$ is a time-dependent deterministic vector field (a drift), 
and $\,v_t(x)\,$ 
is a Gaussian random vector field with mean zero and covariance
\begin{equation}
\big\langle v^{i}_t(x)\,v^{j}_{s}(y)\big\rangle\ =\ \delta(t-s)
\,D_{t}^{ij}(x,y)\,.  \label{dep2}
\end{equation}
Due to the white-noise nature of the temporal dependence of $\,v_t\,$ 
(typical $\,v_t\,$ are distributional in time), \,Eq.\,(\ref{E0}) is
a stochastic differential equation (SDE). We shall consider 
it with the Stratonovich convention\footnote{The choice 
of the Stratonovich convention guarantees that $\,u_t\,$ and $\,v_t\,$ 
transform as vector fields under a change of\\ \hspace*{0.6cm}coordinates.}
\cite{Risk,Oksen}, keeping for the Stratonovich SDEs the notation of the 
ordinary differential equations (ODEs). Examples of systems described by
Eq.\,(\ref{E0}) include four special cases that we shall keep in mind. 
\vskip 0.5cm

\noindent{\bf Example 1.}\ \ Deterministic dynamics
\vskip 0.2cm

\noindent Here $\,v_t(x)\equiv0\,$ and $\,D_t^{ij}(x,y)\equiv 0\,$
so that Eq.\,(\ref{E0}) reduces to the ODE
\qq
\dot{x}\ =\ u_t(x)\,.  
\label{E0det}
\qqq
\vskip 0.4cm

\noindent{\bf Example 2.}\ \ Lagrangian flow in the Kraichnan model
\vskip 0.2cm

\noindent This is a process used in modeling turbulent transport.
The SDE (\ref{E0}), where one usually takes $\,u_t(x)\equiv0$,
describes the motion of tracer particles in a stationary Gaussian 
ensemble of velocities $\,v_t(x)\,$ white in time. Such an ensemble, 
with an appropriate time-independent spatial covariance $\,D^{ij}(x,y)$, 
\,was designed by Kraichnan \cite{Kr68} to mimic turbulent velocities. 
In particular, homogeneous flows are modeled by imposing the translation 
invariance $\,D^{ij}(x,y)=D^{ij}(x-y)\,$ and isotropic ones by assuming 
that $\,D^{ij}(x,y)\,$ is rotation-covariant. In this paper, we shall
consider only the case when $\,D^{ij}(x,y)\,$ is smooth. A discussion 
of the case with $\,D^{ij}(x,y)\,$ non-smooth around the diagonal, 
pertaining to the fully developed turbulence, may be found in \cite{FGV}, 
or, on the mathematical level, in \cite{LeJR}.
\vskip 0.5cm

\noindent{\bf Example 3.}\ \ Langevin dynamics 
\vskip 0.2cm

\noindent Here Eq.\,(\ref{E0}) takes the form\footnote{We use throughout 
the paper the summation convention.}
\qq
\dot{x}^i\,=\,-\,\Gamma^{ij}\partial_jH_t(x)\,
+\,\Pi^{ij}\partial_jH_t(x)\,+\,G_t^i(x)\,+\,\zeta^i(x)\,,
\label{Lang}
\qqq
where $\,\Gamma\,$ is a constant non-negative matrix and $\,\Pi\,$ 
an antisymmetric one, \,the Hamiltonian $\,H_t$ is a, possibly 
time dependent, function, $\,G_t\,$ is an additional force, and 
$\,\zeta_t\,$ is the $\,d$-dimensional white noise with the covariance
\qq
\big\langle\zeta^i_t\,\zeta^j_s\big\rangle\,=\,2\,\delta(t-s)\,
\beta^{-1}\Gamma^{ij}\,.
\label{spind}
\qqq
In this example, the white noise $\,\zeta_t\,$ that plays 
the role of the (space-independent) random vector field $\,v_t\,$ so that 
$\,D_t^{ij}(x,y)=2\beta^{-1}\Gamma^{ij}$. \,For $\,G_t\equiv0\,$ 
and a time independent Hamiltonian $\,H_t\equiv H$, \,the Langevin 
dynamics is used to model the approach to thermal equilibrium at inverse 
temperature $\,\beta$ \cite{HH}. The deterministic vector field $\,-\Gamma^{ij}
\partial_jH\,$ drives the solution towards the minimum of $\,H\,$ 
(if it exists) whereas the Hamiltonian vector field $\,\Pi^{ij}\,
\partial_jH\,$ preserves $\,H$. \,The noise $\,\zeta_t\,$ generates 
the thermal fluctuations of the solution. Note that its spatial covariance
is aligned with the matrix $\,\Gamma\,$ appearing in the dissipative force
$\,-\Gamma^{ij}\partial_jH\,$ (such an alignment, known from Einstein's
theory of Brownian motion, is often called the Einstein relation).
Inclusion of the Hamiltonian vector field permits to model systems where 
the noise acts only on some degrees of freedom, e.g.\ the ones at the ends 
of a coupled chain, with the rest of the degrees of freedom undergoing 
a Hamiltonian dynamics. The introduction of a time-dependence and/or 
of the force $\,G_t\,$ permits to model nonequilibrium systems. 
In the particular case of vanishing $\,\Gamma$, \,the SDE (\ref{Lang}) 
reduces to the ODE
\qq
\dot{x}^i\,=\,\Pi^{ij}\partial_jH_t(x)\,+\,G_t^i(x)
\label{Lang1}
\qqq
describing a deterministic Hamiltonian dynamics in the presence of
an additional force $\,G_t$. 
\eject

\noindent{\bf Example 4.}\ \ Langevin-Kramers equation
\vskip 0.2cm

\noindent This is a special case of the Langevin dynamics that takes place 
in the phase space of $n$ degrees of freedom with $\,x=(q,p)\,$ and 
\qq
\Gamma\,=\,\Big(\begin{matrix}_0&_0\cr^0&^\gamma\end{matrix}\Big)\,,\qquad
\Pi\,=\,\Big(\begin{matrix}_0&_1\cr^{-1}&^0\end{matrix}\Big)\,,\qquad H_t\,=\,
\frac{_1}{^2}\, p\cdot m^{-1}p\,+\,V_t(q)\,,\qquad G_t\,=\,(0,\,f_t(q)),
\nonumber
\qqq
where $\,\gamma\not=0\,$ is a non-negative $n\times n$ matrix, $\,m^{-1}\,$ a
positive one, and $\,1\,$ the unit one. \,Here, Eq.\,(\ref{Lang}) reduces
to the standard relation $\,p_i=m_{ij}\dot{q}^j\,$ between momenta 
and velocities, where $\,m\,$ is the mass matrix, and to the second 
order SDE
\qq
m_{ij}\ddot{q}^j\,=\,-
\gamma_{ik}\dot{q}^k\,-\,\partial_iV_t(q)\,+f_{ti}(q)\,+\,\zeta_i\,,
\label{Kram}
\qqq
that we shall call Langevin-Kramers equation, with the $n$-dimensional 
white noise $\,\zeta\,$ such that
\qq
\big\langle\zeta_{it}\,\zeta_{jt'}\big\rangle\,=\,2\beta^{-1}\gamma_{ij}
\,\delta(t-t')\,.
\nonumber
\qqq
The Langevin-Kramers equation has the form of the Newton equation 
with the friction $\,-\gamma\dot{q}\,$ and white-noise $\,\zeta_t\,$
forces supplementing the conservative one $\,-\nabla V_t\,$ and 
the additional one $\,f_t$. \,It was discussed in \cite{Kurchan1} 
in a very similar context. In the limit of strongly overdamped system 
when the friction term becomes much larger then the 
second order one, the Langevin-Kramers equation (\ref{Kram}) reduces
to the first order SDE
\qq
\gamma_{ik}\dot{q}^k\,=\,-\,\partial_iV_t(q)\,+f_{ti}(q)\,+\,\zeta_i\,,
\nonumber
\qqq
which, if $\,\gamma>0$, \,may be cast again into the form
(\ref{Lang}) but with $\,\Gamma=\gamma^{-1}$, $\,\Pi=0\,$
and $\,H_t=V_t$.
\,One should keep in mind this change when applying the 
results described below for the Langevin dynamics (\ref{Lang})
to the overdamped Langevin-Kramers dynamics.
\vskip 0.3cm

\nsection{Transition probabilities and detailed balance}
\label{sec:trprob}

\noindent Let us recall some basic facts about the diffusion processes
in order to set the notations. We shall denote by $\,\bE^{t_0}_x\,$ 
the expectation of functionals of the Markov process $\,\sfx_t\,$ 
solving the SDE (\ref{E0}) with the initial condition $\,\sfx_{t_0}=x$. 
\,For $t\geq t_0$, \,the relation
\qq
\bE^{t_0}_x\,g(\sfx_{t})\,=\,\int P_{t_0,t}(x,dy)\,g(y)\,
\equiv\,(P_{t_0,t}g)(x)
\label{Ptt}
\qqq
defines the transition probabilities $\,P_{t_0,t}(x,dy)\,$ of the process 
$\,\sfx_t\,$ and the operator $\,P_{t_0,t}$. The transition probabilities 
satisfy the normalization condition $\,\int P_{t_0,t}(x,dy)=1\,$ and the 
Chapman-Kolmogorov chain rule
\qq
\int P_{t_0,t}(x,dy)\,P_{t,t'}(y,dz)\,=\,P_{t_0,t'}(x,dz)\,.
\nonumber
\qqq 
The evolution of the expectation values is governed by 
the second-order differential operators $L_{t}$ defined by the relation
\qq
\frac{d}{dt}\,\bE^{t_0}_x\,g(\sfx_{t})\ =\ 
\bE^{t_0}_x\,(L_{t}g)(\sfx_{t})\,.  
\label{E2}
\qqq
The explicit form of $\,L_t\,$ is found by a standard argument that 
involves the passage from the Stratonovich to the It\^o convention.
For reader's convenience, we give the details in Appendix A. The
result is:
\qq
L_t\,=\,\hat{u}^{i}_t\partial _{i}+\frac{_1}{^2}\partial
_{j}d_{t}^{ij}\partial _{i}\,,  
\label{Generator}
\qqq
where
\qq
d_{t}^{ij}(x)=D_{t}^{ij}(x,x)\qquad{\rm and}\qquad  
\hat{u}^i_t(x)=u^i_t(x)-\frac{_1}{^2}
\partial_{y^j}D_t^{ij}(x,y)|_{y=x}\,.
\label{stand}
\qqq
Due to the relation (\ref{Ptt}), \,Eq.\,(\ref{E2}) may be rewritten as 
the operator identity $\,\partial _{t}P_{t_0,t}=P_{t_0,t}L_{t}$. 
\,Together with the initial condition $\,P_{t_0,t_0}=1$, \,it implies that 
$P_{t_0,t}$ is given by the time-ordered exponential 
\begin{equation}
P_{t_0,t}\,=\,\overrightarrow{\cal T}\,\exp\Big(\int\limits_{t_0}^{t}
L_{s}\,ds\Big)
\,=\,\sum_{n=0}^{\infty
}\,\int\limits_{t_0\leq s_{1}\leq s_{2}\leq ....\leq
t}\hspace{-0.5cm}L_{s_{1}}L_{s_{2}}....L_{s_{n}}\,ds_{1}ds_{2}...ds_{n}\,.
\end{equation}
In particular, $\,P_{t_0,t}=\ee^{(t-t_0)L}\equiv P_{t-t_0}\,$ in the 
stationary case with $\,u_t\equiv u\,$ and $\,D_t\equiv D$.
\,The operator $\,L_t\equiv L\,$ is then called the {\bf generator} of 
the process. 
\vskip 0.3cm

The stochastic process $\,\sfx_t\,$ may be used to evolve measures.
Under the stochastic dynamics, the initial measure $\,\mu_{t_0}(dx)\,$
evolves at time $\,t\,$ to the measure
\qq
\mu_t(dy)\,=\,\int\mu_{t_0}(dx)\,P_{t_0,t}(x,dy)\,.
\label{dynrho}
\qqq
We shall use below the shorthand notation: $\,\mu_t=\mu_0P_{0,t}$. 
\,For measures with densities $\,\mu_t(dx)=\rho_t(x)\,dx\,$ with respect
to the Lebesgue measure $\,dx$, \,Eq.\,(\ref{dynrho}) is equivalent 
to the evolution equation
\qq
\partial_t\rho_t\,=\,\partial_i\big(-\hat u^i_t+\frac{_1}{^2}
d^{ij}_t\partial_j
\big)\rho_t\ \equiv\ L_t^\dagger\rho_t\,,
\label{Ldag}
\qqq
where $\,L^{\dagger }_t\,$ is the (formal) adjoint of the operator $\,L_t$. 
\,The latter relation may be rewritten as the continuity equation
\qq
\partial_t\rho_t\,+\,\nabla\cdot j\,=\,0\qquad{\rm with}\qquad
j^i_t\,=\,(\hat u_{t}^{i}-\frac{_1}{^2}d_{t}^{ij}\partial_{j})\rho_t\,,
\label{prcur}
\qqq 
where $\,\nabla\cdot j\equiv\partial_ij^i_t\,$ is the divergence
of the {\bf density current} $\,j_t\,$ corresponding 
to the measure $\,\mu_t\,$ (the probability current, if $\,\mu_t\,$
is normalized). \,In the case with no explicit time dependence when 
$\,L_t\equiv L$, \,an invariant density $\,\rho$, \,corresponding to an 
invariant measure $\,\mu(dx)=\rho(x)\,dx\,$ of the process, satisfies 
the equation $\,L^{\dagger }\rho=0\,$ which may be rewritten in the 
form of the current conservation condition $\,\nabla\cdot j=0$. \,We 
shall often write the invariant density $\,\rho(x)\,$ in the exponential 
form as $\,\ee^{-\varphi(x)}$. \,One says that the process satisfies 
the {\bf detailed balance relation} with respect to $\,\varphi\,$ if 
the density current $\,j\,$  related to the measure $\,\mu(dx)
=\ee^{-\varphi(x)}dx\,$ vanishes itself, i.e.\ if
\qq
\hat u^i=-\frac{_1}{^2}d^{ij}\partial_j\varphi\,,
\nonumber
\qqq
Equivalently, this condition may be written 
as the relation
\qq
L^\dagger=\ee^{-\varphi}L\,\ee^{\varphi}\,,
\nonumber
\qqq
for the generator of the process or as the identity
\qq
\mu(dx)\,P_t(x,dy)\,=\,\mu(dy)\,P_t(y,dx)
\label{DBP}
\qqq
for the transition probabilities. 
In all these three forms, it implies directly that $\,\mu\,$ 
is an invariant measure. The converse, however, is not true: there exist 
stationary diffusion processes with invariant measures that do not satisfy 
the detailed balance relation.
\vskip 0.2cm

The generator of the stationary Langevin equation
with $\,\Pi=0\,$ and $\,G=0\,$ satisfies the detailed balance
relation with respect to $\,\varphi=\beta H\,$ so that the Gibbs density
$\,\rho(x)=\ee^{-\beta H(x)}$, \,and, if the latter is normalizable, 
the Gibbs probability measure $\,\mu^G(dx)=Z^{-1}\ee^{-\beta H(x)}dx$, 
\,are invariant under such dynamics. The invariance still holds when 
$\,\Pi\not=0\,$ but, in this case, the detailed balance relation fails. 
We shall see below how to generalize the latter to catch also the case 
with conservative forces when $\,\Pi\not=0$.
\vskip 0.3cm

\nsection{Tangent process and phase-space contraction}
\label{sec:tanproc}

One may generate other processes of a similar nature from the diffusive 
process (\ref{E0}). Such constructions will play an important role 
in studying fluctuation relations. As the first example, let us 
consider the separation $\,\delta\sfx_t\,$ between the solution $\,\sfx_t\,$
of Eq.\,(\ref{E0}) with the initial value $\,\sfx_0=x\,$ and another
solution infinitesimally close to $\,\sfx_t$. \,Such a separations evolves 
according to the law
\qq
\delta\sfx_t\ =\ \sfX_t(x)\,\delta\sfx_0\,,
\nonumber
\qqq
where the matrix $\,\sfX_t(x)\,$ with the entries
\qq
\sfX^i_{t\,j}(x)\,=\,\frac{\partial\sfx^i_t}{\partial\sfx_0^j}(x)
\label{entr}
\qqq 
solves the (Stratonovich) SDE
\qq
\dot{X}^i_{\,\,j}\ =\ \big(\partial_k u^i_t
+\partial_k v^i_t\big)(\sfx_t)\,X^k_{\,\,j}
\label{tpr}
\qqq
with the initial condition $\,\sfX_0(x)=1$. \,Together with Eq.\,(\ref{E0}),
the SDE (\ref{tpr}) defines a diffusion process $\,(\sfx_t,\sfX_t)\,$
that we shall call the {\bf tangent process}. In particular, the quantity 
$\,-\ln\det\sfX_t\,$ that represents the accumulated 
phase-space contraction along the trajectory $\,\sfx_t$, \,solves the SDE
\qq
\frac{_d}{^{dt}}\big(-\ln\det X\big)\ =\ -(\nabla\cdot u_t
+\nabla\cdot v_t)(\sfx_t)\,.
\label{phscr}
\qqq
The right hand side of Eq.\,(\ref{phscr}) is the 
{\bf phase-space contraction rate}. We infer that
\qq
-\ln\det\sfX_t\ =\ -\int\limits_0^T(\nabla\cdot u_t)(\sfx_t)\,dt
\,-\,\int\limits_0^T(\nabla\cdot v_t)(\sfx_t)\,dt\,.
\label{phsc}
\qqq
The second integral on the right hand side should be interpreted 
with the Stratonovich convention. The phase-space contraction 
is an important quantity in the study of nonequilibrium dynamics 
and it will reappear in the sequel. 
\vskip 0.3cm

\nsection{Backward processes}
\label{sec:back}

Among the diffusion processes that may be generated from the original process
(\ref{E0}) are the ones which may be interpreted as its time reversals.
The action of time inversion on space-time will be given by the transformation
\qq
(t,x)\ \ \longrightarrow\ \ (T-t,x^*)\,\equiv\,(t^*,x^*)
\label{tmrev}
\qqq
for an involution $\,x\mapsto x^*$. \,It may be lifted to the level
of process trajectories by defining the transformed trajectory 
$\,\tilde\sfx_{t}\,$ by the relation
\qq
\tilde\sfx_{t}\,=\,\sfx_{t^*}^*\,.
\label{tildex}
\qqq
In general, however, we shall {\bf not} define the time-reversed process 
as $\,\tilde\sfx_{t}\,$ because, in the presence of dissipative 
deterministic forces like friction, such time inversion would lead to 
an anti-dissipative dynamics. We shall then allow for more flexibility.
In order to define the time-reversed process, we shall divide the 
deterministic vector field $\,u_t\,$ into two parts 
\qq
u_t\,=\,u_{t,+}+\,u_{t,-}\,,
\label{split}
\qqq
that we shall loosely term dissipative and conservative, choosing
different time-inversion rules for them. The time-reversed  
process $\,\sfx'_{t}\,$ will be given by the SDE
\qq
\dot{x}'\,=\,u'_{t}(x')+v'_{t}(x')
\label{inversion}
\qqq
with the deterministic vector field $\,u'_{t}=u'_{t,+}+\,u'_{t,-}\,$ 
and the random one $\,v'_{t}\,$ defined by the equations
\qq
{u'}^{i}_{t,\pm}(x)
=\pm(\partial_k{x^*}^{i})(x^*)\,u_{t^*,\pm}^{\,k}(x^*)\qquad{\rm and}\ \quad 
{v'}^{i}_{t}(x)=\pm(\partial_k{x^*}^{i})(x^*)\,v^{k}_{t^*}(x^*)\,.
\label{utou'}
\qqq
Note that $\,u_{t,+}\,$ transforms as a vector field under the involution 
$\,x\mapsto x^*\,$ and $\,u_{t,-}\,$ as a pseudo-vector 
field. For $\,v_t\,$ we may use whichever rule since $\,v_t\,$ and 
$\,-v_t\,$ have the same distribution. \,The SDE 
(\ref{inversion}) for the time-reversed process $\,\sfx'_{t}\,$ coincides 
with the one for the process $\,\tilde\sfx_{t}$ defined by 
Eq.\,(\ref{tildex}) if and only if $\,u_{t,+}\,$ vanishes and 
$\,v_t\,$ is transformed according to the pseudo-vector rule. We shall 
call $\,\sfx'_{t}\,$ the {\bf backward process} referring to $\,\sfx_t\,$
as the forward one. The random vector field $\,v'_{t}\,$ of the backward 
process is again Gaussian with mean zero and white-noise behavior in time. 
Its covariance is 
\qq
\big\langle\,v'^{i}_{t}(x)\,v'^{j}_{s}(y)\,\big\rangle\,=\,
\delta(t-s)\,D'^{ij}_{t}(x,y)\,,
\nonumber
\qqq
where
\qq
D'^{ij}_{t}(x,y)\,=\,(\partial_k{x^*}^{i})(x^*)
\,D^{kl}_{t^*}(x^*,y^*)\,(\partial_l{x^*}^{j})(y^*)\,.
\label{D'D}
\qqq
As before, see Eqs.\,(\ref{stand}), we shall denote
\qq
d'^{ij}_{t}(x)\,=\,D'^{ij}_{t}(x,x)\,,\qquad {\hat u}'^{i}_{t}(x)\,
=\,u'^{i}_{t}(x)-\frac{_1}{^2}\partial_{y^{j}}
D'^{ij}_{t}(x,y)|_{y=x}\,.
\label{stand'}
\qqq
\vskip 0.4cm

\noindent{\bf Remark 1.}\ \ Using the chain rule 
$\,(\partial_{j}{x^*}^i)(x^*)(\partial_k{x^*}^{j})(x)=\delta^i_k$, 
\,it is easy to see that the time-inversion transformations 
(\ref{utou'}) are involutive.
\vskip 0.5cm

\noindent Let us emphasize that the choice of a time inversion consists 
of the choice of the involution (\ref{tmrev}) and of the splitting 
(\ref{split}) of $\,u_t$. \,We shall call the process {\bf time-reversible} 
(for a given choice of time inversion) if the deterministic vector fields 
$\,u\,$ and $\,u'\,$ of the forward and of the backward processes coincide 
and if the respective random vector fields $\,v_t\,$ and $\,v'_t\,$ have 
the same distribution, i.e.\ if
\qq
u^i_{t,+}(x)\,+\,u^i_{t,-}(x)\,=\,(\partial_k{x^*}^i)(x^*)\,
\big(u^k_{t^*,+}(x^*)\,-\,u^k_{t^*,-}(x^*)\big)
\nonumber
\qqq
\vskip -0.2cm
\noindent and if
\vskip -0.7cm
\qq
D^{ij}_t(x,y)\,=\,(\partial_k{x^*}^i)(x^*)\,
D^{kl}_{t^*}(x^*,y^*)\,(\partial_l{x^*}^j)(x^*)\,.\label{DDR}
\qqq
Note that the first identity is equivalent to the relations 
\qq
u^i_{t,\pm}(x)\,=\,\frac{_1}{^2}\big(u^i_t(x)\,\pm\,
(\partial_k{x^*}^i)(x^*)\,u^k_{t^*}(x^*)\big)
\label{saspl}
\qqq
and can be always achieved by taking such a splitting of $\,u_t$.
\,It may be not easy, however, to realize  physically the backward 
process corresponding to the splitting (\ref{saspl}). The second 
condition (\ref{DDR}) is a non-trivial constraint on the distribution
of the the white-noise velocity $\,v_t$. \,Nevertheless,
if $\,D_t\,$ is time-independent, \,it may be satisfied by choosing the 
trivial involution $\,x^*\equiv x$.
\vskip 0.4cm

Parallelly to the splitting (\ref{split}) of the drifts 
$\,u_t\,$ and $\,u'_{t}$, \,we shall divide the operators generating
the forward and the backward evolution into two parts:
\qq
L_t\,=\,L_{t,+}+\,L_{t,-}\,,\qquad L'_t=L'_{t,+}+L'_{t,-}
\nonumber
\qqq
according to the formulae:
\qq
&&L_{t,+}\,
=\,\hat u^i_{t,-}\partial_i+\frac{_1}{^2}\partial_jd_t^{ij}\partial_i\,,
\,\hspace{0.02cm}\qquad\quad\,L_{t,-}\,=\,u^i_{t,-}\partial_i\,,\cr
&&L'_{t,+}\,
=\,{\hat{u'}}^{i}_{t,-}\partial_{i}+\frac{_1}{^2}\partial_{j}
d'^{ij}_{t}\partial_{i}\,,\,\,\qquad\ \,L'_{t,-}\,=\,{u'}^{i}_{t,-}
\partial_{i}\,.
\nonumber
\qqq
The time-inversion rules become even more transparent when expressed 
in terms of the split generators. Let $\,R\,$ denote the involution operator 
acting on the functions by
\qq
(Rf)(x)\,=\,f(x^*)\,.
\label{Rop}
\qqq
\vskip 0.3cm

\noindent{\bf Lemma 1.}
\qq
L'_{t,\pm}\,=\,\pm\,R\,L_{t^*,\pm}R\,.
\label{revL}
\qqq
\vskip 0.4cm

\noindent Proof of Lemma 1, involving a straightforward although
somewhat tedious check, is given in Appendix B. 
\vskip 0.4cm

Below, similarly as for 
the forward process, we shall denote by $\,\bE'^{t_0}_{x}\,$ 
the expectation of functionals of the backward process satisfying 
the initial condition $\,\sfx'_{t_0}=x$. \,For $\,t\geq t_0$, 
\,the relations 
\qq
\bE'^{t_0}_{x}\,g(\sfx'_{t})\,=\,(P'_{t_0,t}g)(x)
\qquad
{\rm with}\qquad P'_{t_0,t}\,
=\,\overrightarrow{\cal T}\,\exp\Big(\int\limits_{t_0}^{t}
L'_{s}\,ds\Big)
\nonumber
\qqq
define the operators whose kernels give the transition probabilities 
of the time-reversed process $\,\sfx'_{t}$.
\vskip 0.4cm

\nsection{Examples of time-inversion rules}
\label{sec:extimer}

The preceding considerations were very general. Physically, not all 
time-inversion rules for the diffusive processes (\ref{E0}) described above 
are on the equal footing. In particular situations, some rules may be 
more natural or easier to implement than the other ones. Let us list here 
few cases of special time inversions that were discussed in the literature 
and/or will be used below.

\subsection{Natural time inversion}
\label{sec:nat}

Taking the trivial splitting $\,u_{t,+}=0$, $\,u_{t,-}=u_t\,$
combined with an involution $\,x\mapsto x^*\,$ leads
to the time-inversion rules that produce  the backward process
with trajectories related by the transformation (\ref{tildex})
to the ones of the forward process if the pseudo-vector field rule 
is used when transforming $\,v_t$. \,This is the time inversion
usually employed for the deterministic systems but it may be used
more generally.

\subsection{Time inversion with $\,\,\hat u_{t,+}=0$}
\label{sec:u+=0}

Consider the time inversion corresponding to an arbitrary involution 
$\,x\mapsto x^*\,$ and the choice
\qq
\hat u_{t,+}\,=\,0\,,\qquad u_{t,-}\,=\,\hat u_t\,.
\label{hatrev}
\qqq
of the splitting of $\,u_t$. \,Such a time inversion is a slight 
modification of the natural one to which it reduces
in the case of deterministic dynamics (\ref{E0det}) 
with $\,v_t\equiv 0$. \,As we show in Appendix C, the backward 
dynamics corresponding to the splitting (\ref{hatrev}) is given 
by the relations
\qq
\hat u'^i_{t,+}(x)\,=\,\frac{_1}{^2}d'^{ij}_{t}(x)
\,(\partial_j\ln\sigma)
(x)\,,\qquad u'^i_{t,-}(x)\,=\,-(\partial_k{x^*}^i)(x^*)
\,\hat u^k_{t^*}(x^*)\,,
\label{hatback}
\qqq
where $\,\sigma(x)=\sigma(x^*)^{-1}\,$ denotes the absolute value
$\,|\det(\partial_j{x^*}^{i})(x)|\,$ of the Jacobian of the 
involution $\,x\mapsto x^*$. \,The time inversion considered 
here will be used to obtain fluctuation relations in the 
limiting case of deterministic dynamics (\ref{E0det}) when 
$\,D^{ij}_t\,$ is set to zero and the backward dynamics is 
given by the ODE
\qq
\dot{x}^{\prime i}\,=\,u'^i_{t}(x')\qquad{\rm for}\qquad
u'^i_{t}(x)=-(\partial_k{x^*}^i)(x^*)\,u^k_{t^*}(x^*)\,.
\label{detback}
\qqq
obtained from the ODE (\ref{E0det}) by the natural
time inversion.

\subsection{Time inversion in the Langevin dynamics}
\label{sec:langdyn}

To explain why the rules of time inversion with non-vanishing 
$\,u_{t,+}\,$ are more generally needed, we consider the case 
of the Langevin dynamics that involves the dissipative force
$\,-\Gamma\nabla H_t$. \,Let us arbitrarily split the 
corresponding drift $\,u_t\,$ into two parts:
\qq
u_t\,=\,-\Gamma\nabla H_t\,+\,\Pi\nabla H_t
\,+\,G_t\,=\,u_{t,+}+u_{t,-}\,,
\label{11}
\qqq  
see Eq.\,(\ref{Lang}). Recall the relation (\ref{spind}) that aligns
the matrix $\,\Gamma\,$ with the covariance of the white-noise 
$\,v_t=\zeta_t$. \,It is natural to require the backward dynamics to 
be also of the Langevin type but for the time-reversed Hamiltonian 
$\,H'_{t}(x)=H_{t^*}(x^*)$. \,This requires that
\qq
u'_{t}\,=\,-\Gamma'\nabla H'_{t}\,+\,\Pi'\nabla H'_{t}
\,+\,G'_{t}\,=\,u'_{t,+}+u'_{t,-}\,,
\label{12}
\qqq
and that $\,v'_{t}(x)=\zeta'_{t}\,$ with the covariance of the
white noise $\,\zeta'_{t}\,$ aligned with matrix $\,\Gamma'\,$
as in Eq.\,(\ref{spind}). Upon restriction to linear involutions 
$\,x^*=rx\,$ with $\,r^2=1$, \,the transformation rules (\ref{utou'}) 
become
\qq
u'_{t,\pm}(x)\,=\,\pm r u_{t^*,\pm}(rx)\,,\qquad \zeta'_{t}\,
=\,\pm r\zeta_{t^*}\,.
\nonumber
\qqq
The condition on the covariance of $\,\zeta'_{t}\,$ 
imposes the relation $\,\Gamma'=r\Gamma r^T$. \,Applying $\,r\,$ to the 
both sides of Eq.\,(\ref{12}) taken at time $\,t^*\,$ and at point $\,rx$, 
\,we infer that 
\qq
-r\Gamma'r^T\nabla H_t(x)\,+\,r\Pi'r^T\nabla H_t(x)\,+\,rG'_{t^*}(rx)
\,=\,u_{t,+}(x)-u_{t,-}(x)\,.
\nonumber
\qqq
The latter identity, together with Eq.\,(\ref{11}), result in the relations
\qq
&&u_{t,+}(x)\,=\,-\Gamma\nabla H_t(x)+\frac{_1}{^2}(\Pi+r\Pi'r^T)
\nabla H_t(x)\,+\,\frac{_1}{^2}(G_t(x)+rG'_{t^*}(rx))\,,\cr
&&u_{t,-}(x)\,=\,-\frac{_1}{^2}(\Pi-r\Pi'r^T)\nabla H_t(x)
\,+\,\frac{_1}{^2}(G_t(x)-rG'_{t^*}(rx))\,.
\nonumber
\qqq
At least when $\,\Gamma\,$ is strictly positive, $\,H_t\,$
is not a constant, and the extra force $\,G_t\,$ is absent,
one infers that the component $\,u_{t,+}\,$ cannot vanish identically
by considering the contraction $\,(\nabla H_t)\cdot u_{t,+}$.
\,We shall call {\bf canonical} a choice of the time inversion for
the Langevin dynamics for which 
\qq
&&\Gamma'=r\Gamma r^T=\Gamma\,,\qquad\qquad\,\Pi'=-r\Pi r^T=\Pi\,,\qquad
\label{cano0}\\
&&u_{t,+}\,=\,-\Gamma\nabla H_t\,,\hspace{0.6cm}
\qquad u_{t,-}\,=\,\Pi\nabla H_t\,+\,G_t\,.
\label{cano}
\qqq
Note that such a time inversion treats the force $\,G_t\,$ as a part 
of $\,u_{t,-}\,$ even when this force is of the non-conservative type. 
The Langevin dynamics is time-reversible under a canonical time inversion 
if $\,H'_t=H_t\,$ and $\,G'_t=G_t$. \,For the Langevin-Kramers equation, 
the standard phase-space involution $\,(q,p)^*=r(q,p)=(q,-p)\,$ verifies 
Eqs.\,(\ref{cano0}) and it leads to the particularly simple canonical 
time-inversion rules with
\qq
V'_{t}=V_{t^*}\,,\qquad\quad f'_{t}=f_{t^*}
\nonumber
\qqq
and to the time-reversibility if $\,V_t=V_{t^*}\,$ and $\,f_t=f_{t^*}$.

\subsection{Reversed protocol}
\label{sec:revprot}

The time inversion corresponding to the choice
\qq
u_{t,+}\,=\,u_t\,,\qquad u_{t,-}\,=\,0
\label{revprot}
\qqq
and trivial involution $\,x^*\equiv x\,$ was termed in \cite{ChChJarz} a 
reversed protocol. It may be viewed as consisting of the inversion of 
the time-parametrization in the vector fields in the SDE (\ref{E0}), 
if the vector-field rule is used to reverse $\,v_t$. \,In the stationary case, 
where it results in time-reversibility, such a time inversion was employed 
already in \cite{LebowSp}. \,Here, we shall admit also a possibility 
of a non-trivial involution $\,x\mapsto x^*$. \,The reversed protocol 
leads then to the backward process with 
\qq
u'^i_{t,+}\,=\,(\partial_k{x^*}^i)(x^*)\,u^k_{t^*}(x^*)\,,
\qquad {u'}^i_{t,-}\,=\,0\,,\qquad v'^i_{t}\,=\,(\partial_k{x^*}^i)(x^*)
\,v^k_{t^*}(x^*)\,.
\label{revprot'}
\qqq

\subsection{Current reversal}
\label{sec:currev}

Suppose that $\,\ee^{-\varphi_t}\,$ are densities satisfying 
$\,L_t^\dagger\ee^{-\varphi_t}=0$.  \,Such densities would be  
preserved by the evolution if the generator of the process were frozen 
to $\,L_t$. \,The density current corresponding to $\,\ee^{-\varphi_t}\,$
has the form
\qq
j^i_t\,=\,\big(\hat u^i_t+\frac{_1}{^2}d^{ij}_t(\partial_j\varphi_t)
\big)\,\ee^{-\varphi_t}\,,
\nonumber
\qqq
see Eq.\,(\ref{prcur}). It is conserved due to the relation 
$\,L_t^\dagger\ee^{-\varphi_t}=0$. \,The time inversion defined 
by the choice
\qq
\hat u^i_{t,+}\,=\,-\frac{_1}{^2}d_t^{ij}
\partial_i\varphi_t\,,
\qquad u^i_{t,-}\,=\,\hat u^i_t\,+\,\frac{_1}{^2}d_t^{ij}
\partial_i\varphi_t
\label{CRsplit}
\qqq 
and an arbitrary involution $\,x\mapsto x^*\,$ leads, after an easy 
calculation using the results of Appendix C, to the backward process with
\qq
\hat u'^i_{t,+}\,=\,-\,\frac{_1}{^2}
d'^{ij}_{t}\partial_j\varphi'_{t}\,,\qquad u'^i_{t,-}(x)\,=\,
-(\partial_k{x^*}^i)(x^*)\,u^k_{t^*,-}(x^*)\,
\qquad
v'^i_{t}\,=\,\pm(\partial_k{x^*}^i)(x^*)\,v^i_{t^*}(x^*)
\label{u'CR}
\qqq
for $\,\varphi'_{t}(x)=(\varphi_{t^*}+\ln{\sigma})(x^*)$. \,The density 
current for the backward process corresponding to the densities
$\,\ee^{-\varphi'_{t}}\,$  is
\qq
&\displaystyle{j'^i_{t}\,=\,\big(\hat u'^i_{t}+\frac{_1}{^2}d'^{ij}_{t}
(\partial_j\varphi'_{t})\big)\,\ee^{-\varphi'_{t}}\,
=\,u'^i_{t,-}(x)\,\ee^{-\varphi'_t(x)}\ =\ 
-(\partial_k{x^*}^i)(x^*)\,u^k_{t^*,-}(x^*)\,\ee^{-\varphi_{t^*}(x^*)}
\sigma(x)}\cr
&\displaystyle{=\ -(\partial_k{x^*}^i)(x^*)
\,\big(\hat u^k_{t^*}(x^*)\,+\frac{_1}{^2}d^{ij}_{t^*}(x^*)
(\partial_j\varphi_{t^*})(x^*)\big)\,\ee^{-\varphi_t}\,\sigma(x)
=\ -\,(\partial_k{x^*}^i)(x^*)\,j^k(x^*)\,\sigma(x)}&
\label{CRrev}
\qqq
and is also conserved, as is easy to check. It follows that 
$\,L'^\dagger_{t}\ee^{-\varphi'_{t}}=0$. \,We shall term 
the time inversion corresponding to the choices (\ref{CRsplit})
the {\bf current reversal}. For $\,x^*\equiv x\,$ when it just reverses
the sign of the current, it was already employed in an implicit way  
in \cite{HatSas}, and was introduced explicitly (under a different
name) in \cite{ChChJarz}. \,The latter reference discussed also a simple 
two-dimensional model for which the inverse protocol and the current 
reversal led to different backward processes.

\subsection{Complete reversal}
\label{sec:comprev}

Finally, modifying slightly the last scheme, let us suppose
the densities $\,\rho_t=\ee^{-\varphi_t}\,$ evolve under the dynamics
solving Eq.\,(\ref{Ldag}). With the same splitting
(\ref{CRsplit}) as for the current reversal, \,we obtain the backward 
process for which Eqs.\,(\ref{u'CR}) and (\ref{CRrev}) still hold for
$\,\varphi'_{t}(x)=(\varphi_{t^*}+\ln{\sigma})(x^*)$. \,We shall call 
the corresponding time inversion the {\bf complete reversal}. 
\,Unlike in the other examples, it depends also on the choice of
the initial density $\,\rho_0\,$ and may be difficult to realize 
physically. The time-reflected densities $\,\rho'_t
=\ee^{-\varphi'_t}\,$ evolve now according to the backward-process 
version of Eq.\,(\ref{Ldag}). The current reversal and the complete 
reversal coincide in the case without explicit time dependence and
with the choice of $\,\varphi_t\equiv\varphi\,$ such that 
$\,\ee^{-\varphi}dx\,$ is an invariant measure. 
\vskip 0.3cm

\nsection{Relation between forward and backward processes} 
\label{sec:forwback}

A comparison between the forward and the backward processes
will be at the core of fluctuation relations
that we shall discuss. To put the processes in the two 
time directions back-to-back, we shall adapt to the present
setup the arguments developed in Sect.\,5 of \cite{LebowSp}.
Let us introduce a perturbed version of the generator $\,L_t\,$ 
of the forward process,
\qq
L^1_t\,=\,L_{t}-2\,\hat u^{i}_{t,+}\partial_{i}
-(\partial_{i}\hat u^{i}_{t,+})+(\partial_{i}u^{i}_{t,-})\,.
\label{L1}
\qqq
Operator $\,L^1_t\,$ is related in a simple way to the generator 
of the backward process:
\qq
&R\,\big(L^1_t\big)^\dagger R\,=\,R\,\big(\partial_{i}\hat u^{i}_{t,+}
-\partial_{i}u^{i}_{t,-}+
+\frac{_1}{^2}\partial_{i}d^{ij}_{t}\partial_{j}
-(\partial_{i}\hat u^{i}_{t,+})+(\partial_{i}u^{i}_{t,-})\big)\,R&\cr\cr
&=\,R\,L_{t,+}R\,-\,R\,L_{t,-}R\,=\,L'_{t^*}\,,&
\label{l1l'}
\qqq
where $\,R\,$ is defined by Eq.\,(\ref{Rop}) and the last equality is a 
consequence of the relations (\ref{revL}). Let us consider the time-ordered 
exponential of the integral of $\,L^1_t$.
\,Using the relation $\,L_t^1=(R\hspace{0.025cm}L'_{t^*}R)^\dagger\,$
that follows from Eq.\,(\ref{l1l'}), we infer that
\qq
&&P^1_{t_0,t}\,\equiv\,\overrightarrow{\cal T}\,\exp\Big(
\int\limits_{t_0}^{t}L^1_{s}\,ds\Big)\,=\,
\overleftarrow{\cal T}\,\exp\Big(
\int\limits_{t^*}^{t^*_0}(R\,L'_{s}R)^\dagger\,ds\Big)\cr
&&\qquad=\,\Big[R\,\,\overrightarrow{\cal T}\,\exp\Big(
\int\limits_{t^*}^{t^*_0}\,L'_{s}\,ds\Big)\,R\Big]^\dagger\,=\,
\big(R\,P'_{t^*,t^*_0}R\big)^\dagger\,.
\label{P1P'}
\qqq
Above, the first inversion of the time order from $\,\overrightarrow{\cal T}
\,$ to $\,\overleftarrow{\cal T}\,$ was due to the change 
of integration variables $\,s\mapsto s^*=T-s$, \,and the second one, 
to the fact that the hermitian conjugation reverses the order in 
the product of operators. Let us remark that $\,A(y,dx)\,$ is 
the kernel of the operator $\,A^\dagger\,$ and $\,A(x^*,dy^*)\,$ 
of the operator $\,RAR\,$ if $\,A(x,dy)\,$ is the kernel of a real 
operator $\,A$.
\,Rewriting Eq.\,(\ref{P1P'}) in terms of the kernels, with these 
comments in mind, we obtain the identity
\qq
\,dx\,\,P^1_{t_0,t}(x,dy)\,=\,dy\,\,P'_{t^*,t^*_0}(y^*,dx^*)\,.
\label{P1P'k}
\qqq 
\vskip 0.4cm

\noindent{\bf Remark 2.}\ \ The transition probability of the
backward process on the right hand side may be replaced by the one
of the forward process in the time-reversible case. 
\vskip 0.5cm

Note that the $2^{\rm nd}$ order differential operator $\,L^1_t\,$ 
differs from $\,L_t\,$ only by lower order terms, see Eq.\,(\ref{L1}).
A combination of the Cameron-Martin-Girsanov and the Feynman-Kac formulae 
\cite{StVar} permits to express the kernel $\,P^1_{t_0,t}(x,dy)\,$ as 
a perturbed expectation for the forward process. 
\vskip 0.6cm

\noindent{\bf Lemma 2.}\ \ \ If the matrix $\,\big(d^{ij}_t(x)\big)\,$
is invertible for all $\,t\,$ and $\,x\,$ then
\qq
P^1_{t_0,t}(x,dy)\,=\,\bE^{t_0}_x\ \ee^{-\int\limits_{t_0}^t\CJ_s\,ds}\,
\delta(\sfx_t-y)\,dy\,,
\label{W0}
\qqq
where 
\qq
\CJ_t\,=\,2\,\hat u_{t,+}(\sfx_t)\cdot d_t^{-1}
(\sfx_t)\,\dot{\sfx}_t\,
-\,2\,\hat u_{t,+}(\sfx_t)\cdot d_t^{-1}(\sfx_t)
\,u_{t,-}(\sfx_t)\,-\,(\nabla\cdot u_{t,-})(\sfx_t)
\label{js}
\qqq
is a (local) functional of the solution $\sfx_t$ of the SDE (\ref{E0}).
The right hand side of Eq.\,(\ref{js}) uses the vector notation.
The first term in the expression for $\,\CJ_t\,$ has to be interpreted 
with the Stratonovich convention. 
\vskip 0.6cm

\noindent Proof of Lemma 2 is deferred to Appendix D. \ A combination of
the relations (\ref{W0}) and (\ref{P1P'k}) gives immediately
\vskip 0.8cm

\noindent{\bf Proposition 1}.
\qq
dx\,\,\bE^{t_0}_x\ \ee^{-\int\limits_{t_0}^t\CJ_s\,ds}\,
\delta(\sfx_t-y)\,dy\ =\ dy\,\,P'_{t^*,t^*_0}(y^*,dx^*)\,.
\label{P1P}
\qqq
\vskip -0cm
\hspace{10cm}$\Box$
\vskip 0.6cm

\noindent This is the first fluctuation relation of a series to be 
considered. It connects the transition probability of the 
backward process to an expectation in the forward process weighted
with an exponential factor. Let us illustrate this relation in 
a few particular situations related to the examples of the diffusion
processes considered in Sect.\,\ref{sec:forw}.
\eject

\noindent{\bf Example 5.}\ \ Tangent process in the stationary homogeneous 
Kraichnan model
\vskip 0.2cm

\noindent Recall Sect.\,\ref{sec:tanproc} devoted to the definition
of a tangent process. Let us consider the tangent process 
$\,(\sfx_t,\sfX_t)\,$ with fixed initial data $\,\sfx_0=x\,$ and 
$\,\sfX_0=1\,$ for the homogeneous Kraichnan model. 
As was discussed in detail in \cite{GawWarw}, in this case, the 
distribution of the process $\,\sfX_t\,$ may be obtained by solving, instead 
of the SDE (\ref{tpr}) with $\,u_t\equiv 0$, a simpler
linear It\^o SDE 
\qq
dX\ =\ S_tdt\,X
\label{tprI}
\qqq
with a matrix-valued white-noise $\,S_t\,$ such that
\qq
\big\langle S^i_{t\,k}\,S^j_{s\,l}\big\rangle\ =\ -\delta(t-s)\,
\partial_{k}\partial_{l}D^{ij}(0)\,.
\nonumber
\qqq
In other words, in Eq.\,(\ref{tpr}), we may replace
$\,\partial_kv^i(\sfx_t)\,$ by $\,\partial_kv^i_t(0)\equiv S^i_{t\,k}$,
\,if we change the SDE convention to the It\^o one at the same time.
Consequently, in the homogeneous Kraichnan model, the process 
$\,\sfX_t\,$ may be decoupled from the original process $\,\sfx_t$.
\,Let us abbreviate: $\,-\partial_{k}\partial_{l}D_t^{ij}(0)=C^{ij}_{kl}$.
Remark the symmetries $\,C^{ij}_{kl}=C^{ji}_{lk}=C^{ij}_{lk}=C^{ji}_{kl}$.
\,The It\^o SDE (\ref{tprI}) may be rewritten as the equation
\qq
\dot{X}^i_{\,\,j}\ =\ -\hf C^{ik}_{kl}\,X^l_{\,\,j}\,+\,S^i_{t\,l}
\,X^l_{\,\,j}
\label{tprS}
\qqq
that employs the Stratonovich convention. Upon the use of the notations:
\qq
U^i_{\,\,j}(X)\,=\,-\hf C^{ik}_{kl}\,X^l_{\,\,j}\,,\qquad
V^i_{t\,j}(X)\,=\,S^i_{t\,l}\,X^l_{\,\,j}\,,
\nonumber
\qqq
it may be cast into the form
\qq
\dot{X}\ =\ U(X)\,+\,V_t(X)\,,
\label{E0W}
\qqq
falling within the scope of (stationary) diffusion SDEs 
(\ref{E0}) and defining a Markov process $\,\sfX_t$.
\,The covariance of the white-noise ``velocity'' $\,V_t(X)\,$ is
\qq
\Big\langle V^i_{t\,k}(X)\,V^j_{s\,l}(Y)\Big\rangle\ =\ 
\delta(t-s)\,D^{ij}_{kl}(X,Y)\qquad{\rm with}\qquad 
D^{ij}_{kl}(X,Y)\,=\,C^{ij}_{nm}\,X^n_{\,\,k}\,Y^m_{\,\,l}\,.
\nonumber
\qqq
As in the general case (\ref{stand}), we shall denote:
\qq
d^{ij}_{kl}(X)=D^{ij}_{kl}(X,X)\,,\ \quad
\hat U^i_{\,\,j}(X)=U^i_{\,\,j}(X)-\hf\,
\partial_{\tilde X^k_l}D^{ik}_{jl}(X,Y)\big|_{Y=X}
=-\frac{d+1}{2}C^{ni}_{nk}\,X^k_j\,.\ 
\nonumber
\qqq
\vskip 0.2cm

Let us apply the reversed-protocol time inversion discussed in 
Sect.\,\ref{sec:revprot} to the forward SDE (\ref{E0W}). 
It corresponds to the trivial splitting of $\,U\,$ with $\,U_{+}=U\,$ 
and $\,U_{-}=0\,$ and to an involution $\,X\mapsto X^*\,$
that we shall also take trivial: $\,X^*\equiv X$. \,The backward 
evolution is then given by the same equation (\ref{tprS}) with 
$\,S_t\,$ replaced by $\,S'_t=S_{t^*}$, \,a matrix-valued white noise 
with the same distribution as $\,S_t$. \,The time-reversibility follows. 
Suppose that the covariance $\,C\,$ of the white noise $\,S(t)\,$ 
is invertible\footnote{The assumption about inversibility of $\,C\,$ 
may be dropped at the end by a limiting argument.}, \,i.e.\ that 
there exists a matrix $\,(C^{-1})^{ln}_{jm}\,$ such that 
$\,C^{ij}_{kl}(C^{-1})^{ln}_{jm}=\delta^i_m\delta^n_k$. \,Then
the matrix 
\qq
(d^{-1})^{ln}_{jm}(X)\,=\,(X^{-1})^l_{\,\,p}(X^{-1})^n_{\,\,r}
(C^{-1})^{pr}_{jm}
\nonumber
\qqq
provides the inverse of $\,d^{ij}_{kl}(X)$.
\,Substituting these data into Eq.\,(\ref{js}), we obtain
\qq
\CJ_t=2(\hat U)^j_{\,\,l}(\sfX_t)\,(d^{-1})^{ln}_{jm}(\sfX_t)\,
\dot{\sfX}^m_{t\ n}
=-(d+1)\,(\sfX_t^{-1})^n_{\,\,m}\,\dot{\sfX}^m_{t\ n}
=-(d+1)\,\frac{d}{dt}\ln|\det\sfX_t|.\ \ 
\nonumber
\qqq
The relation (\ref{P1P}) applied to the case at hand leads to the identity
\qq
dX_0\,\,P_{t}(X_0,dX)\,|\det X_0|^{-(d+1)}|\det X|^{d+1}\ =\ 
dX\,\,P_{t}(X,dX_0)\,,
\label{KrFR0}
\qqq
where $\,P_t(X_0,dX)\,$ denotes the transition probability
of the forward process $\,\sfX_t\,$ solving 
the SDEs (\ref{tprI}) or (\ref{tprS}) and $\,dX_0\,$ on 
the left hand side and $\,dX\,$ on the right hand side stand for the
Lebesgue measures on the space of $\,d\times d\,$ matrices.
We made use of the fact that the backward process has the same 
law as the forward one. Eq.\,(\ref{KrFR0}) is nothing else
a the detailed balance relation with respect to 
$\,\varphi(X)=\ln|\det X|$. \,Indeed, note that the density
current corresponding to the density $\,\rho(X)=|\det X|^{-d+1}\,$
\qq
j(X)\ =\ \big(\hat U(X)+\frac{_1}{^2}d(X)\nabla\varphi\big)\,|\det X|^{-d+1}\,,
\nonumber
\qqq
see Eq.\,(\ref{prcur}), vanishes. 
\vskip 0.2cm

Integrating the left hand side of the
above identity against a function $\,f(X_0,X)\,$ and
using the relation $\,P_t(X_0,dX) =P_t(1,d(XX_0^{-1}))\,$ that 
follows from the invariance of the corresponding SDE under the 
right multiplication of $\,X\,$ by invertible matrices, we
obtain the equalities
\qq
&\displaystyle{\int f(X_0,X)\,\det(XX_0^{-1})^{d+1}\,dX_0\,P_t(X_0,dX)\,=
\int f(X_0,X)\,\det(XX_0^{-1})^{d+1}\,dX_0\,P_t(1,d(XX_0^{-1}))}&\cr
&\displaystyle{=\int f(X_0,XX_0)\,(\det X)^{d+1}\,dX_0\,P_t(1,dX)\,
=\int f(X^{-1}X_0,X_0)\,(\det X)\,dX_0\,P_t(1,dX)\,,}&
\nonumber
\qqq
where we twice changed variables in the iterated integrals.
On the other hand, the integration of the right hand side of Eq.\,(\ref{KrFR0})
against $\,f(X_0,X)\,$ gives 
\qq
&\displaystyle{\int f(X_0,X)\,dX\,P_t(X,dX_0)\,
=\int f(X,X_0)\,dX_0\,P_t(X_0,dX)
\,=\int f(XX_0,X_0)\,dX_0\,P_t(1,dX)}&\cr
&\displaystyle{=\int f(X^{-1}X_0,X_0)\,dX_0\,P_t(1,dX^{-1})\,.}&
\nonumber
\qqq  
Comparing the two expressions, we infer that
\qq
P_t(1,dX)\,(\det X)\ =\ P_t(1,dX^{-1})\,.
\label{KrFR}
\qqq
This is a version of the Evans-Searles \cite{EvSear} fluctuation 
relation for the stationary homogeneous Kraichnan model. In the 
context of general hydrodynamic flows, it was formulated and proven 
by a change-of-integration-variables argument in \cite{BFF}, 
see also \cite{GawWarw}. We shall return in Sect.\,\ref{sec:stationFR}
to the relation (\ref{KrFR}) in order to examine some of its consequences. 
Subsequently, we shall generalize it in Sect.\,\ref{sec:MFR} to arbitrary 
diffusion processes of the type (\ref{E0}). 
\vskip 0.7cm

\noindent{\bf Example 6.}\ \ Generalized detailed balance relation
\vskip 0.2cm

\noindent Consider the complete-reversal rules discussed
in Sect.\,\ref{sec:comprev} and corresponding to the choice (\ref{CRsplit}).
Since, by virtue of the assumption that the densities $\,\ee^{-\varphi_t}\,$ 
evolve under the dynamics, see Eq.\,(\ref{Ldag}),
\qq
L_t^\dagger\ee^{-\varphi_t}\,=\,-\partial_i\,u^i_{t,-}
\ee^{-\varphi_t}\,=\,\ee^{-\varphi_t}\big(u^i_{t,-}
\partial_i\varphi_t-\partial_iu^i_{t,-}\big)\,=\,-
\ee^{-\varphi_t(x)}\,(\partial_t\varphi)(x)\,,
\label{varphit}
\qqq
the last two terms in the definition (\ref{js}) 
reduce to $-\,(\partial_t\varphi_t)(\sfx_t)\,$ in this case so that
\qq
\CJ_t\ =\ -(\nabla\varphi_t)(\sfx_t)\cdot\dot{\sfx}_t\,
-\,(\partial_t\varphi_t)(\sfx_t)\ =\ -\frac{d}{dt}\,\varphi_t(\sfx_t)\,.
\label{hk}
\qqq
Upon integration over time, this produces boundary terms and
Eq.\,(\ref{P1P}) implies the
\,{\bf generalized detailed balance relation}:
\qq
\mu_{0}(dx)\,P_{0,T}(x,dy)\,=\,\mu_T(dy)\,
P'_{0,T}(y^*,dx^*)\,,
\label{DB1}
\qqq
for $\,\mu_t(dx)=\ee^{-\varphi_t(x)}dx$.  
\,Note that Eq.\,(\ref{DB1}) holds for any choice of the involution 
$\,x\mapsto x^*$. \,Upon integration over $\,x$, \,Eq.\,(\ref{DB1}) 
assures that the measures $\,\mu_t\,$ stay
invariant under the dynamics, what was assumed from the very beginning.
In the case with no explicit time dependence, i.e.\ when $\,L_t\equiv L$,
\,Eq.\,(\ref{DB1}) holds, in particular, for $\,\varphi_t\equiv\varphi\,$
such that $\,\mu=\ee^{-\varphi}dx\,$ is an invariant measure. \,In that case, 
the generalized detailed balance relation reduces to the detailed balance one 
(\ref{DBP}) if $\,u_{-}\,$ in the splitting (\ref{CRsplit}) vanishes and 
$\,x^*\equiv x$. \,This was the case in Example 5. Below, we shall see 
examples where the invariant measure $\,\mu\,$ is known and the 
generalized detailed balance relation holds but where the detailed 
balance itself fails. Some of those cases fall under the scope 
of the Langevin dynamics. Let us discuss them first. 
\vskip 0.7cm

\noindent{\bf Example 7.}\ \ $1^{\rm st}$ law of thermodynamics 
and generalized detailed balance for the Langevin dynamics 
\vskip 0.2cm

\noindent For the Langevin dynamics with the splitting 
(\ref{cano}) of the drift, a direct substitution yields
\qq
\CJ_t&=&-\beta(\nabla H_t)(\sfx_t)
\cdot\dot{\sfx}_t+\beta(\nabla H)(\sfx_t)\cdot G_t(\sfx_t)
-(\nabla\cdot G_t)(\sfx_t)\,.
\nonumber
\qqq
Upon the use of the dynamical equation (\ref{Lang}), 
\qq
\int\limits_{0}^T\CJ_t\,dt\,= 
\int\limits_{0}^T\big[\beta\,(\nabla H_t)(\sfx_t)\cdot
\Gamma(\nabla H_t)(\sfx_t)\,-\,
\beta\,(\nabla H_t)(\sfx_t)\cdot\zeta_t\,-\,(\nabla\cdot G_t)(\sfx_t)\big]dt
\ \equiv\ \beta\sfQ\,,
\label{bQ}
\qqq
where  $\,\sfQ\,$ may be identified with the heat
transfered to the environment modeled by the thermal noise. 
On the other hand, using the original expression for $\,\CJ_t\,$ 
together with the (Stranonovich convention) identity 
$\,\frac{d}{dt}\,H_t(\sfx_t)=(\nabla H_t)(\sfx_t)
\cdot\dot{\sfx}_t+(\partial_t H_t)(\sfx_t)$, \,we obtain the relation
\qq
\int\limits_{0}^T\CJ_t\,dt\ =\ -\beta\Delta\sfU\,+\,\beta\sfW
\,,
\label{LangeW}
\qqq
where $\,\Delta\sfU=H_T(\sfx_T)-H_{0}(\sfx_{0})\,$ is
the change of the internal energy of the system and
\qq
\sfW\ =\ \int\limits_{0}^T
\big[(\partial_tH_t)(\sfx_t)+(\nabla H)(\sfx_t)\cdot G_t(\sfx_t)
-\beta^{-1}(\nabla\cdot G_t)(\sfx_t)\big]dt
\label{work}
\qqq  
may be interpreted as the work performed on the system. With this
interpretations, a comparison of the two expressions for the integral 
of $\,\CJ_t\,$ leads to the $\,{\bf 1^{\rm st}}\,$ {\bf law of 
thermodynamics}:
\qq
\Delta\sfU\ =\ -\sfQ\ +\ \sfW\,.
\qqq
This was discussed in a simple example of the forced and damped oscillator 
in \cite{JGC}. In the absence of the extra force $\,G_t$, \,the expression
for the work reduces to
\qq
\sfW\ =\ \int\limits_{0}^T(\partial_tH_t)(\sfx_t)\,dt
\label{JarzW}
\qqq
and represents the so called {\bf Jarzynski work} introduced
first in \cite{Jarz1} for deterministic Hamiltonian dynamics. 
In the stochastic Langevin-Kramers dynamics, the expressions for 
the heat and the work become:
\qq
\sfQ\ =\ \int\limits_{0}^T\big[\dot{q}_t\cdot\gamma\,\dot{q}_t\,
-\,\dot{q}_t\cdot\zeta_t\big]dt\,,\qquad
\sfW\ =\ \int\limits_{0}^T
\big[(\partial_tV_t)(q_t)\,+\,\dot{q}_t\cdot f_t(q_t)
\big]dt\,.
\label{QW}
\qqq
The second quantity is equal to the sum of the Jarzynski work and 
of the work of the external force $\,f_t$. \,It was introduced
and discussed in \cite{Kurchan1}. In the stationary case, it reduces 
to the injected work \cite{Kurchan} and, up to the $\beta$-factor, 
coincides with the ``action functional'' (for uniform temperature) 
given by Eq.\,(6.3) of \cite{LebowSp}. \,Note that the general 
expression (\ref{work}) for work makes also sense in the case 
of deterministic dynamics (\ref{Lang1}) obtained from the SDE 
(\ref{Lang}) by setting $\,\Gamma=0$, \,in particular for the 
deterministic Hamiltonian evolution with $\,G_t\equiv0$. 
\vskip 0.2cm

If $\,G_t\equiv0$, \,the splitting (\ref{cano}) is a special
case of the splitting used for the current reversal for 
$\,\varphi_t=\beta H_t$, \,see Eq.\,(\ref{CRsplit}).
In particular, if $\,H_t\equiv H\,$ then the transition 
probabilities of the Langevin process satisfy the generalized
detailed balance relation (\ref{DB1}) that takes the form
\qq
\mu(dx)\,P_T(x,dy)\ =\ \mu(dy)\,P'_T(y^*,dx^*)
\label{GDBP}
\qqq
for $\,\mu(dx)=\ee^{-\beta H(x)}dx\,$ and any involution 
$\,x\mapsto x^*=rx$.
\,The latter identity replaces the detailed balance relation 
(\ref{DBP}) in the presence of the conservative force 
$\,\Pi\nabla H\,$ and still assures that the Gibbs density 
$\,\ee^{-\beta H}\,$ is invariant 
under such Langevin dynamics. If the involution $\,r\,$ satisfies 
additionally the relations (\ref{cano0}) and $\,H(rx)=H(x)$, 
\,resulting in the time-reversibility, then one may replace 
$\,P'_T\,$ by $\,P_T\,$ in the relation (\ref{DB1}). 
\vskip 0.7cm

\noindent{\bf Example 8.}\ \ Linear Langevin equation
\vskip 0.2cm

\noindent Consider the linear SDE
\qq
\dot{x}\ =\ M\,x\,+\,\zeta_t\,,  
\label{E0L}
\qqq
where $\,M\,$ is a $\,d\times d\,$ matrix and $\,\zeta_t\,$ is the white
noise with the covariance (\ref{spind}) and matrix $\,\Gamma\,$ strictly
positive. We shall be interested in cases when the matrix $\,\Gamma^{-1}M\,$ 
is non-symmetric. \,For an elementary discussion of mathematical aspects
of such SDEs see e.g. \cite{Hashm}. In the context of nonequilibrium
statistical mechanics, examples of such linear equations were considered 
in \cite{LebowSp0} as models of a harmonic chain of oscillators 
interacting with environment of variable temperature or, 
quite recently, in \cite{TCCP} for modeling coiled polymers in 
a shearing flow. \,The diffusion process $\,\sfx_t\,$ that solves 
Eq.\,(\ref{E0L}) with the initial value $\,\sfx_0=x\,$ is given by 
the formula
\qq
\sfx_t\,=\,\ee^{\,tM}x\,+\,\int\limits_0^t\ee^{\,(t-s)M}\,\zeta_s\,ds\,.
\nonumber
\qqq
The transition probabilities of this process are Gaussian and have
the explicit form
\qq
P_{t}(x,dy)\,=\,\det(2\pi\beta^{-1}C_{t})^{-1/2}\,\,
\exp\Big[-\frac{\beta}{2}\big(y-\ee^{\,tM}x\big)\cdot
C_{t}^{-1}\big(y-\ee^{\,tM}x\big)\Big]dy\,,\ \ 
\label{trproblin}
\qqq
\vskip -0.01cm
\noindent where 
\vskip -0.75cm
\qq
C_{t}\,=\,2\int\limits_{0}^t\ee^{sM}\Gamma\ee^{sM^T}\,ds
\label{Ct}
\qqq
is a strictly positive matrix. 
Suppose that all the eigenvalues $\,\lambda\,$ of $\,M\,$ have
negative real parts. Under this condition, $\,\ee^{tM}\,$ tends
to zero exponentially fast when $\,t\to\infty\,$ so that
$\,C_\infty\equiv C\,$ is finite
and 
\qq
P_{t}(x,dy)\ \ \ \mathop{\longrightarrow}_{t\to\infty}\ \ \ 
\det(2\pi\beta^{-1}C)^{-1/2}
\,\,\exp\Big[-\frac{\beta}{2}\,y\cdot
C^{-1}y\Big]dy\,,
\nonumber
\qqq
with the right hand side defining the unique invariant probability measure
of the process. This Gaussian measure has the form of the Gibbs measure 
for the quadratic Hamiltonian 
\qq 
H(x)\,=\,\frac{_1}{^2}\,x\cdot C^{-1}x\,.
\label{Ham}
\qqq
Introducing the matrix
\qq
\Pi\,=\,\Gamma\,+\,MC\,.
\label{Pi}
\qqq
that is an antisymmetric:
\qq
\Pi\,+\,\Pi^T\,=\,2\Gamma\,+\,2\int\limits_0^\infty\ee^{\,sM}(M\Gamma+\Gamma
M^T)\ee^{\,sM^T}\,ds\,=\,2\Gamma\,+\,2\int\limits_0^\infty
\frac{d}{ds}\,\ee^{\,sM}\Gamma\ee^{\,sM^T}\,ds\,=\,0\,,
\nonumber
\qqq
the linear SDE (\ref{E0L}) may be rewritten in the Langevin
form (\ref{Lang}) as
\qq
\dot{x}\ =\ -\Gamma\nabla H(x)\,+\,\Pi\nabla H(x)\,+\,\zeta_t\,.
\label{E0L1}
\qqq
Conversely, the last SDE with $\,H\,$
as in Eq.\,(\ref{Ham}) for some $\,C>0\,$ is turned  into the form (\ref{E0L})
upon setting
\qq
M\,=\,-(\Gamma-\Pi)C^{-1}\,.
\label{M}
\qqq
Note that the last equation implies the relation (\ref{Pi}) for 
$\,\Pi$. \,In Appendix E, we show that $\,M\,$ given by Eq.\,(\ref{M}) 
has necessarily all eigenvalues with negative real part and that $\,C\,$
may be recovered from $\,M\,$ as $\,C_\infty\,$ 
given by Eq.\,(\ref{Ct}) with $\,t=\infty$. \,This establishes
the equivalence between the SDEs (\ref{E0L}) and (\ref{E0L1}).   
\vskip 0.2cm

The probability current associated by the formula (\ref{prcur}) 
to the Gaussian invariant Gibbs measure 
$\,\mu^G(dx)=Z^{-1}\ee^{-\beta H(x)}dx\,$ is
\qq
j(x)\,=\,Z^{-1}\Pi C^{-1}x\,\ee^{-\beta H(x)}\,.
\nonumber
\qqq
It vanishes only when $\,\Pi=0$.
\,In the latter case, the transition probabilities (\ref{trproblin}) 
satisfy the detailed balance relation (\ref{DBP}) for $\,\varphi=\beta H
+\ln{Z}$. \,If $\,\Pi\not=0\,$ then only a generalized detailed balance
relation (\ref{GDBP}) holds for any choice of the linear 
involution $\,x\mapsto x^*=rx$. \,If moreover $\,r\Gamma r^T=\Gamma$, 
$\,r\Pi r^T=-\Pi\,$ and $\,rC r^T=C\,$ then $\,P'_T\,$ on the right hand 
side of Eq.\,(\ref{DB1}) may be replaced by $\,P_T$.
\vskip 0.3cm

\nsection{Jarzynski equality}
\label{sec:Jarz}

We shall exploit further consequences of the relation (\ref{P1P}) 
between the forward and backward processes. 
In this section we shall derive an identity that generalizes the celebrated 
Jarzynski equality \cite{Jarz1,Jarz2} and shall prepare the ground 
for obtaining more refined fluctuation relations following the ideas 
of \cite{Gall1}, \cite{Maes} and \cite{Crooks2}. 
\,Let $\,\varphi_0\,$ and $\,\varphi_T\,$ be two functions generating 
measures \qq
\mu_0(dx)\,=\,\ee^{-\varphi_0(x)}dx\,,\qquad\mu_T(dx)\,
=\,\ee^{-\varphi_T(x)}dx\,,
\label{measur}
\qqq 
respectively. In particular, we could take
$\,\ee^{-\varphi_T(x)}\,$ such that the measure $\,\mu_T\,$ is related
to $\,\mu_0\,$ by the dynamical evolution
(\ref{dynrho}), i.e. $\,\mu_T=\mu_0P_{0,T}$, \,but we shall not assume 
such a choice unless explicitly stated. In general, the measures 
(\ref{measur}) may be not normalizable but we shall impose the normalization 
condition later on. We shall associate to $\,\mu_0\,$ and $\,\mu_T\,$ 
the time-reflected measures 
\qq
\mu'_0(dx)\,=\,\ee^{-\varphi'_0(x)}dx\,=\,\ee^{-\varphi_T(x^*)}dx^*\,,\qquad
\mu'_T(dx)\,=\,\ee^{-\varphi'_T(x)}dx\,=\,\ee^{-\varphi_0(x^*)}dx^*\,.
\nonumber
\qqq 
Let us modify the functional $\,\int\limits_0^T\CJ_t\,dt\,$ 
introduced in the last section by boundary terms $\,\Delta\varphi\equiv
\varphi_T(\sfx_T)-\varphi_0(\sfx_0)\,$ by setting
\qq
\CW\,=\,\Delta\varphi\,+
\int\limits_0^T\CJ_t\,dt\,.
\label{W0T}
\qqq
The functional $\,\CW\,$ will be the basic quantity in what 
follows. Its physical interpretation in terms of the entropy production 
will be discussed in the Sect.\,\ref{sec:entrprod} below.
\vskip 0.4cm

For any functional $\,\CF\,$ on the space of trajectories $\,\sfx_t\,$ 
parametrized by time in the interval $\,[0,T]$, \,we shall denote by 
$\,\tilde\CF\,$ the functional defined by $\,\tilde\CF(\sfx)=\CF(\tilde\sfx)$, 
\,where $\,\tilde\sfx\,$ is given by Eq.\,(\ref{tildex}). 
We shall also introduce the shorthand notation
\qq
\bE_{x,y}^{0,T}\,\,\CF(\sfx)\,=\,\bE_x^{0}\,\,\CF(\sfx)\,\delta(\sfx_T-y)
\nonumber
\qqq
for the (unnormalized) expectation of the process $\,\sfx_t\,$ with 
fixed initial and final points, and similarly for the backward process.  
The following refinement of the relation (\ref{P1P}) of Proposition 1 holds: 
\vskip 0.8cm

\noindent{\bf Proposition 2}.
\qq
\mu_0(dx)\,\,\bE_{x,y}^{0,T}\,\,\CF(\sfx)\,\,\ee^{-\CW(\sfx)}
\,\,dy\
=\ \mu'_0(dy^*)\,\,\bE'^{\hspace{0.01cm}0,T}_{
y^*\hspace{-0.1cm},x^*}\,\,\tilde\CF(\sfx')\,\,dx^*
\label{BI0}
\qqq
\vskip 0.7cm

\noindent Proof of Proposition 2 is contained in Appendix F. Note that
the explicit dependence on the choice of measures $\,\mu_0\,$ and 
$\,\mu'_0\,$ trivially cancels the one buried in $\,\CW$. \,In particular, 
for $\,\CF\equiv 1$, \,Proposition 2 reduces to Proposition 1 with
$\,t_0=0\,$ and $\,t=T$. \,As before, the backward-process 
expectation $\,\bE'\,$ may be replaced by the forward-process one $\,\bE\,$
for the time-reversible process.  
\vskip 0.6cm

If the measures $\,\mu_0\,$ and $\,\mu'_0\,$
are normalized then we may use them
as the probability distributions of the initial points of the forward 
and of the backward process, respectively. The corresponding probability 
measures $\,M(d\sfx)\,$ and $\,M'(d\sfx')\,$ on the space of trajectories
on the time-interval $\,[0,T]\,$ are given by the relations
\qq
&&\int\CF(\sfx)\,M(d\sfx)\quad\,\,\ =\ \int\Big(\bE_x^0\,\CF(\sfx)\Big)
\,\mu_0(dx)\quad\ \equiv
\ \Big\langle\CF\Big\rangle\,,\label{M(dx)}\\
&&\int\CF(\sfx')\,M'(d\sfx')\ \ \,=\  
\int\Big(\bE'^0_{x}\,\CF(\sfx')\Big)\,\mu'_0(dx)\,\ 
\,\,\equiv\ \,\Big\langle\CF\Big\rangle^{\hspace{-0.06cm}\prime}\,.
\label{M'(dx')}
\qqq
Upon integration over $\,x\,$ and $\,y$, the identity (\ref{BI0})
induces the following equality between the expectations
with respect to the trajectory measures $\,M\,$ and $\,M'$:
\vskip 0.8cm

\noindent{\bf Corollary 1.} 
\qq
\Big\langle\,\CF\,\,\ee^{-\CW}\,
\Big\rangle\ 
=\ \Big\langle\,\tilde\CF\,
\Big\rangle^{\hspace{-0.06cm}\prime}
\label{BI}
\qqq
\vskip 0.6cm

\noindent  It was stressed in \cite{Maes}, and even more 
explicitly in \cite{Crooks2}, that the identity of the type 
of (\ref{BI}), comparing the expectations in the forward and 
the backward processes, is a source of fluctuation relations. 
An important special case of Eq.\,(\ref{BI}) is obtained by setting 
$\,\CF\equiv1$. \,It was derived in \cite{Jarz1} in the context 
of the Hamiltonian dynamics and in \cite{Jarz2} in the one 
of Markov processes:
\vskip 0.7cm

\noindent{\bf Corollary 2 \,(Jarzynski equality).}
\qq
\Big\langle\,\ee^{-\CW}\,\Big\rangle\ =\ 1\,.
\label{Jarz}
\qqq
\vskip 0.7cm

\noindent Let us illustrate the meaning of the above relation 
by considering a few special cases.
\vskip 0.7cm

\noindent{\bf Example 9.}\ \ The case of Langevin dynamics 
\vskip 0.2cm

\noindent With the splitting (\ref{cano}) used for the canonical time 
inversion, upon taking $\,\varphi_t=\beta(H_t-F_t)$, where 
$\,F_t=-\beta^{-1}\ln\int\ee^{-\beta H_t(x)}dx\,$ denotes the  
free energy, \,we infer from Eq.\,(\ref{LangeW}) that
\qq
\CW\ =\ \beta(\sfW\,-\,\Delta F)\,,
\label{JarzW0} 
\qqq
where $\,\Delta F=F_T-F_0\,$ is the free energy change and $\,\sfW\,$ 
is the work given by Eq.\,(\ref{work}). \,The difference $\,\sfW-\Delta F\,$ 
is often called the {\bf dissipative work}. The Jarzynski equality 
(\ref{Jarz}) may be rewritten in this case in the original form 
\qq
\Big\langle\,\ee^{-\beta\sfW}\,\Big\rangle\ =\ \ee^{-\beta\Delta F}\,.
\label{Jarz1}
\qqq
in which it has become a tool to compute the differences between 
free energies of equilibrium states from nonequilibrium processes 
\cite{HJarz,CRJSTB,Ritort, Ritort1}. 
\vskip 0.7cm

\noindent{\bf Example 10.}\ \ The case of deterministic dynamics
\vskip 0.2cm

\noindent Upon splitting the drift $\,u_t\,$ as in Eq.\,(\ref{hatrev}) 
of Sect.\,\ref{sec:u+=0},
the expression (\ref{js}) reduces to $\,\CJ_t=-(\nabla\cdot
\hat u_t)(\sfx_t)$. \,For the deterministic dynamics
where $\,D_t^{ij}(x,y)\equiv0$, \, one has $\,\hat u_t=u_t\,$ so that
\qq
\CJ_t\ =\ -(\nabla\cdot u_t)(\sfx_t)\,.
\label{CJdet}
\qqq
The right hand side represents the phase-space contraction rate
along the trajectory $\,\sfx_t$, \,see Eq.\,(\ref{phscr}). \,In this case,
\qq
\CW\ =\ \Delta\varphi\,-\,
\int\limits_0^T (\nabla\cdot u_{t})(\sfx_t)\,dt\,=\,
\int\limits_0^T\big[\frac{_d}{^{dt}}\varphi_t(\sfx_t)-
(\nabla\cdot u_{t})(\sfx_t)\big]dt\,.
\label{df}
\qqq
\,For $\,\varphi_T=\varphi_0=\varphi$, \,the last integral
in Eq.\,(\ref{df}) was termed ``the integral of the 
{\bf dissipation function}'' in \cite{EvSear2}. In the case of the 
deterministic dynamics (\ref{Lang1}) obtained from the Langevin 
equation by setting $\,\Gamma=0$, \,the expression (\ref{df}) 
for $\,\CW\,$ reduces to the one of Eq.\,(\ref{JarzW0}) if we take 
$\,\varphi_t=\beta(H_t-F_t)$. \,In the deterministic 
case, the Jarzynski equality (\ref{Jarz}) reads
\qq
\int\ee^{\,\int\limits_0^T(\nabla\cdot u_{t})(\sfx_t)\,dt}\,
\ee^{-\varphi_T(\sfx_T)}\,d\sfx_0\ =\ 1
\label{Jarzdet}
\qqq
and  may be easily proven directly. To this end recall Eq.\,(\ref{phsc})
which implies for the deterministic case that $\ \int\limits_0^T
(\nabla\cdot u_{t})(\sfx_t)\,dt=\ln\det\sfX_T(\sfx_0)$,
\ were the matrices $\,\sfX_t(x)\,$ of the tangent process
are given by Eq.\,(\ref{entr}). \,The equality (\ref{Jarzdet}) is then 
obtained by the change of integration variables $\,\sfx_0\mapsto\sfx_T\,$ 
whose Jacobian is equal to $\,\det\sfX_T(\sfx_0)$. 
\vskip 0.7cm

\noindent{\bf Example 11.}\ \ The reversed protocol case
\vskip 0.2cm

\noindent In the setup of Sect.\,\ref{sec:revprot} with $\,u_{t,-}=0$,
\qq
\CJ_t\,=\,2\,\hat u_{t}(\sfx_t)\cdot d_t^{-1}(\sfx_t)\,\dot{\sfx}_t\,.
\nonumber
\qqq
In the stationary case,  the integral $\,\int\limits_0^T\CJ_t\,dt$, 
\,rewritten with use of the It\^o convention, was termed an ``action'' 
in \cite{LebowSp}, see Eq.\,(5.3) therein. In \cite{HatSas}, it was 
considered in the context of the Langevin equation with the extra force 
$\,G_t\,$ (but without the Hamiltonian term $\,\Pi\nabla H_t$). It
was then identified as $\,\beta\sfQ^{tot}\,$ with the quantity 
$\,\sfQ^{tot}\,$  interpreted, following \cite{OP}, as the {\bf total heat} 
produced in the environment. \,The functional $\,\CW\,$ of the forward 
process is given here by the formula
\qq
\CW\ =\ \Delta\varphi\,+\,
2\int\limits_0^T\hat u_t(\sfx_t)\cdot d_t^{-1}(\sfx_t)\,
\dot{\sfx}_t\,dt\ \equiv\ \CW^{tot}\,.
\label{revprotW}
\qqq
In particular, for the Langevin dynamics (\ref{Lang}), \,one obtains: 
\qq
\CW^{tot}\ =\ \Delta\varphi\,+\,\beta\int\limits_0^T
(-\Gamma\nabla H_t+\Pi\nabla H_t+G_t)(\sfx_t)\cdot\Gamma^{-1}
\dot{\sfx}_t\,dt\,.
\label{WtotL}
\qqq
The Jarzynski equality (\ref{Jarz}) was discussed for this case
in \cite{HatSas,SpS,ChChJarz}. 
Note that $\,\CW^{tot}\,$ is not well defined for the Langevin-Kramers 
dynamics. On the other hand, for the linear Langevin equation 
of Example 8 and for $\,\varphi_t=\beta(H-F)$,
\qq
\CW^{tot}\ =\ \beta\int\limits_0^T\sfx_t\cdot(C^{-1}+M^T\Gamma^{-1})
\,\dot{\sfx}_t\,dt\ =\ 
-\beta\int\limits_0^T\sfx_t\cdot C^{-1}\Pi\Gamma^{-1}\dot{\sfx}_t\,dt
\label{Wtotlin}
\qqq
and it vanishes if $\,\Pi=0$. \,A long time asymptotics of the
probability distribution of a quantity differing from the last one 
by a boundary term was studied in \cite{TCCP}.
\vskip 0.7cm

\noindent{\bf Example 12.}\ \ Hatano-Sasa equality \cite{HatSas} 
\vskip 0.2cm

\noindent In the current-reversal setup of Sect.\,\ref{sec:currev}, with 
the splitting (\ref{CRsplit}) of the drift $\,u_t\,$ induced by 
the normalized densities $\,\ee^{-\varphi_t}\,$ such that 
$\,L_t^\dagger\ee^{-\varphi_t}=0$, 
\qq
\CJ_t\ =\ -(\nabla\varphi_t)(\sfx_t)\cdot\dot{\sfx}_t
\label{jsH}
\qqq
since now the last two terms on the right hand side of Eq.\,(\ref{js})
vanish, compare to Eq.\,(\ref{varphit}). Upon integration, this gives:
\qq
\int\limits_0^T\CJ_t\,dt\ =\ -\Delta\varphi\ 
+\ \int\limits_0^T(\partial_t\varphi_t)(\sfx_t)\,dt
\label{jsHS}
\qqq
In \cite{HatSas}, the integral of $\,\CJ_t\,$ given by Eq.\,(\ref{jsH})
was identified in the context of the Langevin equation with the force
$\,G_t\,$ as equal to $\,\beta\sfQ^{ex}\,$ \,where $\,\sfQ^{ex}\,$
was termed the {\bf excess heat}, following \cite{OP}. \,The difference 
$\,\sfQ^{tot}-\sfQ^{ex}=\sfQ^{hk}\,$ was called, in turn, the 
{\bf housekeeping heat} and was interpreted as the heat production 
needed to keep the system in a nonequilibrium stationary state, 
see again \cite{OP,HatSas,SpS,ChChJarz}. Using in the definition 
(\ref{W0T}) the functions $\,\varphi_0\,$ and $\,\varphi_T\,$ from 
the same family, we infer from Eq.\,(\ref{jsHS}) that 
\qq
\CW\ =\ \int\limits_0^T(\partial_t\varphi_t)(\sfx_t)\,dt\ \equiv
\ \CW^{ex}\,.
\label{HSW}
\qqq
The equality (\ref{Jarz}) for this case was proven by
Hatano-Saso \cite{HatSas}, see also \cite{Kurchan1}. 
Note that in the stationary case, $\,\CW^{ex}=0$.
\,The Langevin dynamics discussed in Example 9 provides a special 
instance of the situation considered here if $\,G_t\equiv0$.
\,Consequently, in that case, $\,\CW^{ex}\,$ is equal to 
the dissipative Jarzynski work (in the $\beta^{-1}$ units) 
$\,\beta(\sfW-\Delta F)\,$ with $\,\sfW\,$ given by Eq.\,(\ref{JarzW}).
\vskip 0.7cm

\noindent{\bf Example 13.}\ \ The case of complete reversal 
\vskip 0.2cm

\noindent Recall that for the complete reversal rule of 
Sect.\,\ref{sec:comprev} based on the choice of densities
$\,\ee^{-\varphi_t}\,$ evolving dynamically, $\,\CJ_t\,$ is the total 
time derivative, see Eq.\,(\ref{hk}). The use in the definition (\ref{W0T}) 
of the functions from the same family annihilates the functional
$\,\CW$:
\qq
\CW\ \equiv 0\,.
\label{Wcomprev}
\qqq
\vskip 0.3cm

\nsection{Speck-Seifert equality}
\label{sec:SpSeif}

Let us consider the two functionals $\,\CW^{tot}\,$ and $\,\CW^{ex}\,$ of 
the process $\,\sfx_t\,$ introduced in Examples 11 and 12. 
We shall take them with the same functions $\,\varphi_t\,$ satisfying 
$\,L_t^\dagger\ee^{-\varphi_t}=0$. \,The two Jarzynski 
equalities $\,\Big\langle\ee^{-\CW^{tot}}\Big\rangle
=1=\Big\langle\ee^{-\CW^{ex}}\Big\rangle\,$ hold simultaneously.
In \cite{SpS} a third equality of the same type, this time involving   
the quantity
\qq
\CW^{hk}\ =\ \CW^{tot}-\CW^{ex}\ =\ 
\int\limits_0^T\big[\nabla\varphi_t(\sfx_t)+
2\,\hat u_t(\sfx_t)\cdot d_t^{-1}(\sfx_t)\big]
\dot{\sfx}_t\,dt 
\nonumber
\qqq
was established in the context of the Langevin equation where
$\,\CW^{hk}=\beta\sfQ^{hk}=\beta\sfQ^{tot}-\beta\sfQ^{ex}\,$ 
is the housekeeping heat (in the $\beta^{-1}$ units).
We shall prove here a generalization of the result of \cite{SpS}.
To this end, let us consider, besides the original process $\,\sfx_t\,$
satisfying the SDE (\ref{E0}), \,the Markov process 
$\,\sfx''_t\,$ satisfying the same equation but with the drift 
$\,\hat u_t\,$ replaced by 
\qq
\hat u''_t\,=\,-\hat u_t\,-\,d_t\nabla\varphi_t\,.
\label{u''}
\qqq
We shall denote by $\,\big\langle\ \cdot\ \big\rangle^{\hspace{-0.06cm}
\prime\prime}\,$ the expectation defined by Eq.\,(\ref{M(dx)}) but 
referring to the process $\,\sfx''_t$. \,Note in passing the relations 
$\,L''^\dagger_t\ee^{-\varphi_t}=0$, \,where the operators 
$\,L''_t\,$ are given 
by Eq.\,(\ref{Generator}) with $\,\hat u''_t\,$ replacing $\,\hat u_t$. 
\,In particular, in the stationary case, the processes $\,\sfx_t\,$
and $\,\sfx''_t\,$ have the same invariant measure.
\vskip 0.6cm

\noindent{\bf Proposition 3}.
\qq
\Big\langle\,\CF\,\,\ee^{-\CW^{hk}}\,\Big\rangle\ =\ 
\Big\langle\,\CF\,\Big\rangle^{\hspace{-0.06cm}\prime\prime}.
\label{SpSgen}
\qqq
\vskip 0.6cm

\noindent{\bf Proof.} \ The above identity may be proven directly 
with the use of the Cameron-Martin-Girsanov formula, see Appendix D, 
by comparing the measures of the processes $\,\sfx_t\,$ and 
$\,\sfx''_t\,$ corresponding to SDEs differing by a drift term. 
Here we shall give another proof based on applying twice the relation  
(\ref{BI}). First, we use this relation with the functional 
$\,\CF\,$ replaced by $\,\CF\,\ee^{-\CW^{tot}+2\CW^{ex}}\,$ for the 
current-reversal time inversion with the trivial involution 
$\,x^*\equiv x\,$ and the vector-field rule for $\,v_t$.
\,This results in the equality
\qq
\Big\langle\,\CF\,\ee^{-\CW^{hk}}\,\Big\rangle\ =\ 
\Big\langle\,\tilde\CF\,\ee^{-\tilde\CW^{tot}
+2\tilde\CW^{ex}}
\,\Big\rangle^{\hspace{-0.06cm}\prime}\,,
\label{tored}
\qqq
where the expectation $\,\big\langle\ \cdot\ \big\rangle^{\hspace{-0.06cm}
\prime}\,$ pertains to the backward dynamics with
\qq
\hat u'^i_{t}\,=\,-\hat u^i_{t^*}\,-\,d^{ij}_{t^*}\partial_j\varphi_{t^*}
\,,\qquad v'^i_{t}\,=\,v^i_{t^*}\,,
\nonumber
\qqq
see Eqs.\,(\ref{u'CR}). Now, we observe that the same backward process
may be obtained by the reversed-protocol time inversion, again for
$\,x^*\equiv x$, \,from the process
$\,\sfx''_t\,$ introduced above. \,The identity (\ref{BI}) applied for 
the processes $\,\sfx''_t\,$ and $\,\sfx'_t\,$ reads:
\qq
\Big\langle\,\CF''\,\ee^{-\CW''}\,
\Big\rangle^{\hspace{-0.06cm}\prime\prime}\ =\ 
\Big\langle\,\tilde\CF''\,\Big\rangle^{\hspace{-0.06cm}\prime}\,,
\label{sfx'sfx''}
\qqq
\vskip -0.3cm
\noindent where
\vskip -0.7cm
\qq
\CW''\,=\,\Delta\varphi\,+\,
2\int\limits_0^T\hat u''_t(\sfx_t)\cdot d_t^{-1}(\sfx_t)\,\dot{\sfx}_t\,dt
\nonumber
\qqq
is the functional $\,\CW\,$ referring to the dynamics with
$\,\hat u''_{t}=\hat u''_{t,+}\,$ given by Eq.\,(\ref{u''}). 
The application of Eq.\,(\ref{sfx'sfx''}) to 
$\,\CF''=\CF\,\ee^{-\CW^{tot}+2\CW^{ex}}\,$ reduces the right 
hand side of 
Eq.\,(\ref{tored}) to the expectation $\,\big\langle
\CF\,\ee^{-(\CW^{tot}-2\CW^{ex}+\CW'')}
\big\rangle^{\hspace{-0.06cm}\prime\prime}$. 
\,The equality (\ref{SpSgen}) follows by checking that 
\qq
&\displaystyle{\CW^{tot}-\,2\CW^{ex}+\,\CW''\ =\ 
\Delta\varphi\,+\,
2\int\limits_0^T\hat u_t(\sfx_t)\cdot d_t^{-1}(\sfx_t)\,
\dot{\sfx}_t\,dt\,-\,2\int\limits_0^T(\partial_t\varphi_t)
(\sfx_t)\,dt}&
\nonumber
\qqq
\vskip -1.1cm
\qq
&\displaystyle{+\ \Delta\varphi\,+\,2\int\limits_0^T
\big(-\hat u_t(\sfx_t)-d_t(\sfx_t)\nabla\varphi_t(\sfx_t)\big)
\cdot d_t^{-1}(\sfx_t)\,\dot{\sfx}_t\,dt}&
\nonumber
\qqq
\vskip -1.1cm
\qq
&\displaystyle{=\ 2\Delta\varphi\,-\,
2\int\limits_0^T\big[\partial_t\varphi_t(\sfx_t)
+\nabla\varphi_t(\sfx_t)\cdot{\sfx}_t\big]\,dt\ =\ 0\,.}& 
\nonumber
\qqq
\vskip -0.2cm
\hspace{10cm}$\Box$
\vskip 0.7cm

\noindent Setting $\,\CF\equiv1\,$ in the identity (\ref{SpSgen}), we 
obtain the result that was established by a different argument in \cite{SpS} 
in the context of the Langevin equation:
\vskip 0.7cm

\noindent{\bf Corollary 3 \,(Speck-Seifert equality).}
\qq
\Big\langle\,\ee^{-\CW^{hk}}\,\Big\rangle\ =\ 1\,.
\label{SpSeif}
\qqq
\vskip 0.3cm

\nsection{Entropy production}
\label{sec:entrprod}

An immediate consequence of the Jarzynski equality (\ref{Jarz}) and 
of the Jensen inequality (i.e.\ of convexity of the exponential function) is
\vskip 0.7cm

\noindent{\bf Corollary 4 \,($2^{\rm nd}$ law of thermodynamics for 
diffusion processes).}
\qq
\big\langle \CW\big\rangle\ \geq\ 0\,.
\label{2law}
\qqq
\vskip 0.6cm

\noindent To discuss the relation of the latter inequality to the
$2^{\rm nd}$ law of thermodynamics, let us first remark that
the quantity on the left hand side has the interpretation 
of a relative entropy. Recall, that for two probability measures
$\,\mu(d\sfx)\,$ and $\,\nu(d\sfx)=\ee^{-w(\sfx)}\mu(d\sfx)$,
\,the relative entropy of $\,\nu\,$ with respect to $\,\mu\,$ 
is defined by the formula
\qq
S(\mu|\nu)\ =\ \int w(\sfx)\,\mu(d\sfx)
\nonumber
\qqq 
and is always non-negative. Now, the identity (\ref{BI})
may be read as the relation 
\qq
\tilde M'(d\sfx)\ =\ \ee^{-\CW}\,M(d\sfx)\,.
\nonumber
\qqq 
between the measures $\,M(d\sfx)\,$ and 
$\,\tilde M'(d\sfx)\equiv M'(d\tilde\sfx)$.
In other words, $\,\ee^{-\CW}\,$ is the relative 
(Radon-Nikodym) density of the trajectory measure $\tilde M'$ with 
respect to the measure $\,M$. \,It follows that
\qq
\big\langle \CW\big\rangle\ =\ S(M|\tilde M')
\nonumber
\qqq
so that the inequality (\ref{2law}) expresses the positivity of the 
relative entropy. 
\vskip 0.3cm

Up to now, the measures
$\,\mu_0\,$ and $\,\mu_T\,$ were unrelated.
Let us consider the particular case when $\,\mu_T\,$ is 
obtained by the dynamical evolution (\ref{dynrho}) from
$\,\mu_0\,$ so that $\,\mu_T=\mu_0P_{0,T}$. \,In this case, 
the relative entropy $\,S(M|\tilde M')\,$ may be interpreted 
as the {\bf entropy production} in the forward process between 
times $\,0\,$ and $\,T$, \,relative to the backward process, see 
\cite{EyiLebSp,MaesN,Gasp}. Let for a measure $\,\nu(dx)=\rho(x)dx$, 
$\,S(\nu)=-\int\ln\rho(x)\,\nu(dx)\,$ denotes its entropy.
Using the definition (\ref{W0T}), we may rewrite 
\qq
\big\langle \CW\big\rangle\ =\ 
S(\mu_T)\,-\,S(\mu_0)\,+\,\Delta S_{env}
\label{avW}
\qqq
where the difference $\,S(\mu_T)-S(\mu_0)\,$ is 
the change of entropy of the fixed-time distribution of the process during
the time $\,T\,$ and 
\qq
\Delta S_{env}\ =\ \int\limits_0^T\big\langle\CJ_t\big\rangle\,dt\,.
\nonumber
\qqq
The latter quantity will be interpreted as the mean entropy 
production in the environment modeled by the stochastic noise,
measured relative to the backward process. The quantity 
$\,\big\langle\CJ_t\big\rangle\,$ represents then the instantaneous
mean rate of the entropy production in the environement. 
The inequality (\ref{2law}) states then that the overall entropy 
production cannot be negative in mean. In this sense, it is a version 
of the $2^{\rm nd}$ law of thermodynamics for the diffusion processes 
under consideration. In the stationary case, where $\,\mu_T=\mu_0$, 
\,the overall mean entropy production reduces to the one in the 
environment $\,\Delta S_{env}$. 
\vskip 0.2cm

Note that $\,\Delta S_{env}\,$ defined above depends on the time 
inversion employed (more precisely, on the splitting of $\,u_t$), 
and the quantities obtained by employing different time inversions 
are, in general, different. They may have different physical 
relevance. For the Langevin equation with the splitting 
(\ref{cano}), $\,\Delta S^{env}=\beta\langle\sfQ\rangle$, \,where 
$\,\sfQ\,$ is the heat transfered to the environment given by 
Eq.\,(\ref{bQ}). \,We may talk about the total mean entropy production 
in the environment $\,\Delta S_{env}^{tot}\,$ if the reversed 
protocol of Sect.\,\ref{sec:revprot} and Example 11 is used or about 
the excess mean entropy production $\,\Delta S_{env}^{ex}\,$ in the 
environment for the current reversal of Sect.\,\ref{sec:currev} 
and Example 12. The Speck-Seifert equality (\ref{SpSeif}) combined 
with the Jensen inequality implies that the former does not exceeds 
the latter. As an illustration, consider the stationary Langevin 
equation with vanishing additional force where 
$\,\Delta S^{ex}_{env}=0\,$ although $\,\Delta S^{tot}_{env}\,$ may 
be non-zero if $\,\Pi\not=0$. \,In particular, in the linear case 
studied in Example 8, $\,\CW^{tot}\,$ is given by Eq.\,(\ref{Wtotlin})
and
\qq
\Delta S^{tot}_{env}\ =\ \big\langle\CW^{tot}\big\rangle\ =\ 
-\beta\int\limits_0^T
\big\langle\sfx_t
\cdot C^{-1}\Pi\Gamma^{-1}M{\sfx}_t\big\rangle\,dt\,=\,
-T\,{\rm tr}\,\Pi\Gamma^{-1}M\,,
\nonumber
\qqq
where the second equality was obtained using the SDE (\ref{E0L})
together with the fact that, for the integral 
$\ \int\limits_0^T\sfx_t\cdot C^{-1}\Pi\Gamma^{-1}
\zeta_t\,dt$, \ the Stratonovich and the It\^{o} conventions 
coincide so that its expectation vanishes. \,Finally, note that 
for the complete reversal, the overall entropy production vanishes 
because $\,\CW\equiv0\,$ in this case, see Eq.\,(\ref{Wcomprev}).
With our flexibility of the choice of the backward process, there
are always ones with respect to which there is no entropy 
production!
\vskip 0.3cm

In the deterministic case when $\,\CJ_t\,$ is given by Eq.\,(\ref{CJdet}),
the mean rate of entropy production in the environment is
\qq
\big\langle\CJ_t\big\rangle\,=\,-\int(\nabla\cdot u_t)(x)\,\mu_t(dx)
\nonumber
\qqq
where $\,\mu_t\,$ is obtained by the dynamical evolution from
the measure $\,\mu_0$, \,i.e. $\mu_t=\mu_0P_{0,t}\,$
for $\,P_{0,t}(\sfx_0,dy)=\delta(y-\sfx_t)dy$. 
\,For uniformly hyperbolic dynamical systems without explicit
time dependence, the measures $\,\mu_t\,$ tend for large $\,t\,$ to the 
invariant SRB measure $\,\mu_\infty\,$ and the mean rate of entropy 
production in the environment converges to the expectation of the phase-space 
contraction rate $\,-\nabla\cdot u\,$ with respect to $\,\mu_\infty\,$ 
\cite{Ruelle1}. A discussion of the relation between of the phase-space 
contraction to the production of thermodynamic entropy in deterministic 
dynamics employing models of finite-dimensional thermostats may be found 
in \cite{Gall3}.
\vskip 0.4cm

If the measure $\,\mu_T\,$ is not obtained by evolving dynamically
$\,\mu_0\,$ then one has to distinguish between the measures
$\,\mu_T\,$ and $\,\mu_0P_{0,T}$.
\,In this case, the relation (\ref{avW}) is modified to
\qq
\big\langle \CW\big\rangle\ =\ 
S(\mu_0P_{0,T})\,-\,S(\mu_0)\,
+\,\Delta S_{env}\,+\,S(\mu_0P_{0,T})|\mu_T)\,,
\nonumber
\qqq
i.e. the left hand side is increased by the relative entropy
of the measure $\,\mu_T\,$ with respect 
to the measure obtained from $\,\mu_0\,$ by the dynamical evolution. 
Consequently, the average $\,\big\langle \CW\big\rangle\,$ 
is minimal when $\,\mu_T=\mu_0P_{0,T}$. 
\vskip 0.3cm

\nsection{Linear response for the Langevin dynamics}
\label{sec:LR}
\subsection{Green-Kubo formula and Onsager reciprocity}

\noindent As noted in \cite{ECM,Gall2,LebowSp}, fluctuations 
relations may be viewed as extensions to the non-perturbative 
regime of the Green-Kubo and Onsager relations for the 
nonequilibrium transport coefficients valid within the linear 
response description of the vicinity of the equilibrium. Here, 
for the sake of completeness, we shall show how such relations 
follow formally from the Jarzynski equality (\ref{Jarz}) for the 
Langevin dynamics. To this end, we shall consider the latter with 
a time independent Hamiltonian $\,H_t\equiv H\,$ and the additional 
time-dependent force
\qq
G_t(x)\ =\ g_{ta}\,G^a(x)\,,
\nonumber
\qqq
where the couplings $\,g_{ta}, \,a=1,2$, \,are arbitrary (regular) functions 
of time (and the summation over the index $\,a\,$ is understood). In the 
case at hand, we infer from Eq.\,(\ref{JarzW0}) that 
\qq
\CW\ =\ \int\limits_0^Tg_{ta}\,J^a(\sfx_t)\,dt
\qquad{\rm for}\qquad
J^a\ =\ \beta(\nabla H)\cdot G^{a}-\nabla\cdot G^{a}\,.
\nonumber
\qqq
In particular, for the Langevin-Kramers equation 
(\ref{Kram}), 
\qq
J^a\,=\,\beta\,f^a_i(q)\,\dot{q}^i
\nonumber
\qqq
is the power injected by the external force $\,f^a\,$ 
(in the $\,\beta^{-1}\,$ units). The quantities $\,J^a\,$ are often 
called {\bf fluxes} associated to the forces $\,G^a$.
\vskip 0.3cm
 
Let us denote by $\,\big\langle\CF\big\rangle\,$ the
expectation defined by Eq.\,(\ref{M(dx)}) with $\,\mu_0\,$ standing
for the Gibbs measure $\,Z^{-1}\ee^{-\beta H}dx\,$ and by 
$\,\big\langle\CF\big\rangle_{\hspace{-0.04cm}0}\,$ the same 
expectation taken for $\,g_{ta}\equiv 0$, \,i.e. in the equilibrium
system. \,Expanding Eq.\,(\ref{Jarz}) up to the 
second order in $\,g_{ta}\,$ and abbreviating $\,J^a(\sfx_t)\equiv J^a_t$, 
\,we obtain the identity
\qq
-\int\limits_0^T g_{ta}\,\big\langle J^a_t
\big\rangle_{\hspace{-0.04cm}0}\,dt
\,-\int\limits_0^T\int\limits_0^T g_{ta}\,g_{tb}
\,\big\langle J^a_t\,\CR^{b}_{t'}\big\rangle_{\hspace{-0.04cm}0}\,dt\,dt'
\,+\,\frac{_1}{^2}\int\limits_0^T\int\limits_0^T g_{ta}
\,g_{tb}\,\big\langle J^a_t\,J^b_{t'}\big\rangle_{\hspace{-0.04cm}0}
\,dt\,dt'\ =\ 0\,,
\label{2ord}
\qqq
where the insertion of the response field $\,\CR^a_t\,$ is defined 
by the relation
\qq
\big\langle \CF\,\CR^{a}_t\big\rangle_{\hspace{-0.04cm}0}\ 
=\ \frac{\delta}{\delta g_{ta}}\Big|_{_{g\equiv0}}\big\langle 
\,\CF\,\big\rangle.
\nonumber
\qqq
Note that $\,\big\langle J^a_t\,\CR^{b}_{t'}\big\rangle_0=0\,$
for $\,t'>t\,$ because of the causal nature of the stochastic evolution.
\,The vanishing of the term linear in $\,g_{ta}\,$ in Eq.\,(\ref{2ord})
implies that the equilibrium expectation of the fluxes $\,J^a\,$
vanishes
\qq
\big\langle J^a_t\big\rangle_{\hspace{-0.04cm}0}\ =\ Z_0^{-1}\int J^a(x)\,
\ee^{-\beta\,H_0(x)}\,dx\ =\ 0\,,
\nonumber
\qqq
which is easy to check directly.
\,Stripping the quadratic term in Eq.\,(\ref{2ord}) of arbitrary
functions $\,g_{ta}$, \,we infer that 
\qq
\big\langle J^a_t\,\CR^{b}_{t'}\big\rangle_{\hspace{-0.04cm}0}
\ =\ \theta(t-t')\,\big\langle J^a_t\,
J^b_{t'}\big\rangle_{\hspace{-0.04cm}0}\,.
\nonumber
\qqq
The integration of the latter equation over $\,t'\geq0\,$ results
in the relation
\qq
\frac{_\partial}{^{\partial g_b}}\Big|_{_{g=0}}\big\langle 
J^a_t\big\rangle\ =\ \int\limits_0^t\big\langle J^a_t\,
J^b_{t'}\big\rangle_{\hspace{-0.04cm}0}\,dt'\,,
\label{GK0}
\qqq
where on the left hand side we consider the derivative with
respect to the coupling $\,g_b\,$ constant in time. \,In the limit 
$\,t\to\infty$, \,we may expect the convergence
of the expectation $\,\big\langle J^a_t\big\rangle \,$ 
in the presence of the time-independent force $\,g_aG^a\,$
(and of its derivatives over $\,g_b$) to the nonequilibrium 
stationary expectation $\,\big\langle J^a_t\big\rangle_{\hspace{-0.08cm}st}\,$
(and its derivatives). \,Let us also assume that the temporal decay 
of the stationary equilibrium correlation function of the fluxes 
is sufficiently fast, e.g.\, exponential. These may be often established
for the dynamics governed by the Langevin equation by studying
the properties of its generator. With these assumptions, 
Eq.\,(\ref{GK0}) implies
\vskip0.7cm

\noindent{\bf Proposition 4} \ ({\bf Green-Kubo formula}). 
\vskip -0.38cm
\qq
\frac{_\partial}{^{\partial g_b}}\Big|_{_{g=0}}\big\langle 
J^a_t\big\rangle_{\hspace{-0.08cm}st}\ 
=\ \int\limits_{-\infty}^t\big\langle J^a_t
\,J^b_{t'}\big\rangle_{\hspace{-0.04cm}0}\,dt'\,.
\nonumber
\qqq
\vskip 0.6cm

\noindent The stationary equilibrium correlation function
$\,\big\langle J^a_t\,J^b_{t'}\big\rangle_{\hspace{-0.04cm}0}\,$ depends 
only on the difference $\,t-t'\,$ of times. Besides, if the system is 
time-reversible, then $\,\big\langle J^a_t\,J^b_{t'}
\big\rangle_{\hspace{-0.04cm}0}=
\big\langle J^b_t\,J^a_{t'}\big\rangle_{\hspace{-0.04cm}0}\,$
and the Green-Kubo formula may be rewritten in the form
\qq
\frac{_\partial}{^{\partial g_b}}\Big|_{_{g=0}}\big\langle 
J^a_t\big\rangle_{\hspace{-0.08cm}st}\ =\ \frac{_1}{^2}\int
\big\langle J^a_t\,J^b_{t'}\big\rangle_{\hspace{-0.04cm}0}\,dt'\ 
=\ \frac{_1}{^2}\int\big\langle J^b_t\,J^a_{t'}\big\rangle_{\hspace{-0.04cm}0}
\,dt'
\nonumber
\qqq
which implies 
\vskip 0.8cm

\noindent{\bf Corollary 5 \ (Onsager reciprocity).}
\vskip -0.38cm
\qq
\frac{_\partial}{^{\partial g_b}}\Big|_{_{g=0}}\big\langle 
J^a_t\big\rangle_{\hspace{-0.08cm}st}\ 
=\ \frac{_\partial}{^{\partial g_a}}\Big|_{_{g=0}}
\big\langle J^b_t\big\rangle_{\hspace{-0.08cm}st}\,.
\nonumber
\qqq
\vskip 0.5cm

\subsection{Fluctuation-dissipation theorem}
\label{sec:FDT}

Let us consider again the Jarzynski equality for the Langevin dynamics,
this time in the absence of the additional force $\,G_t\,$ but 
with a time dependent Hamiltonian
\qq
H_t(x)\ =\ H(x)\,-\,h_{ta}\,O^a(x)\,,
\label{Ht}
\qqq
where $\,h_{ta}, \,a=1,2$, \,vanish at $\,t=0\,$ and $\,O^a(x)\,$ 
are functions of $\,x\,$ (``observables''). In this case,
Eq.\,(\ref{JarzW0}) reduces to the relation 
\qq
\CW\ =\ -\beta\int\limits_0^T{\dot h}_{ta}\,O^a(\sfx_t)\,-\,
\beta\Delta F\,,
\nonumber
\qqq
where
\qq
\beta\Delta F\ =\ -\ln\int\ee^{-\beta\big(H(x)\,-\,h_{Ta}O^a(x)\big)}dx\ 
+\ \ln\int\ee^{-\beta H(x)}dx\,.
\nonumber
\qqq
Expanding the left hand side of the Jarzynski equality (\ref{Jarz}) up 
to the second order in $\,h_{ta}\,$ and abbreviating $\,O^a(\sfx_t)
\equiv O^a_t$, \,we infer that
\qq
&\beta\int\limits_0^T\dot{h}_{ta}\big\langle O^a_t
\big\rangle_{\hspace{-0.04cm}0}\,dt\ -\ 
\beta\,h_{Ta}\,\big\langle O^a_0
\big\rangle_{\hspace{-0.04cm}0}\ =\ 0&
\label{1ord}
\qqq
and that
\qq
&\frac{_1}{^2}\beta^2\int\limits_0^T\int\limits_0^T\dot{h}_{ta}\,
\dot{h}_{t'b}\,\big\langle O^a_t\,O^b_{t'}
\big\rangle_{\hspace{-0.04cm}0}\,dt\,dt'\,
+\,\beta\int\limits_0^T\int\limits_0^T\dot{h}_{ta}\,h_{t'b}\,\big\langle
O^a_t\,R^b_{t'}\big\rangle_{\hspace{-0.04cm}0}\,dt\,dt'&\cr\cr  
&-\,\frac{_1}{^2}\beta^2
\,h_{Ta}\,h_{Tb}\,\big\langle O^a_0\,O^b_0
\big\rangle_{\hspace{-0.04cm}0}\ =\ 0\,,&
\label{2orh}
\qqq
where the insertion of the response field $\,R^a_t\,$ 
is defined similarly as that of $\,\CR^a_t\,$ before by
$$\big\langle \CF\,R^a_t\big\rangle_{\hspace{-0.04cm}0}\ 
=\ \frac{\delta}{\delta h_{ta}}
\Big|_{_{h\equiv0}}\big\langle\,\CF\,\big\rangle\,.$$
Again, similarly as before, $\,\big\langle O^a_t\,R^b_{t'}
\big\rangle_{\hspace{-0.04cm}0}=0\,$ for $\,t'>t\,$
because of causality.
\vskip 0.2cm

The first order equality (\ref{1ord}) is equivalent
to the time-independence of the equilibrium expectation of
$\,O^a_t$. \,As for the second order relation (\ref{2orh}),
upon expressing $\,h_{ta}\,$ as the integral of $\,\dot{h}_{ta}$, 
\,it is turned into the equality
\qq
&\beta\int\limits_0^T\int\limits_0^T\dot{h}_{ta}\,
\dot{h}_{t'b}\,\big\langle \left(O^a_t\,O^b_t\,-\,
O^a_t\,O^b_{t'}\right)\big\rangle_{\hspace{-0.04cm}0}\,dt\,dt'
\ =\ 2\int\limits_0^T\int\limits_0^T\int\limits_0^{t'}
\dot{h}_{ta}\,\dot{h}_{t''b}\,\big\langle O^a_t\,R^b_{t'}
\big\rangle_{\hspace{-0.04cm}0}\,
dt\,dt'\,dt''\,.\ \ &
\nonumber
\qqq
After the change of the order of integration over $\,t'\,$ and $\,t''\,$
followed by the interchange of those symbols, the right hand side
becomes 
\qq
&2\int\limits_0^T\int\limits_0^T\int\limits_{t'}^T
\dot{h}_{ta}\,\dot{h}_{t'b}\,\big\langle O^a_t\,R^b_{t''}
\big\rangle_{\hspace{-0.04cm}0}\,
dt\,dt'\,dt''\ 
=\ 2\int\limits_0^T\int\limits_0^T\int\limits_{t'}^t
\dot{h}_{ta}\,\dot{h}_{t'b}\,\theta(t-t')\,\big\langle O^a_t
\,R^b_{t''}\big\rangle_{\hspace{-0.04cm}0}\,
dt\,dt'\,dt''\ \ &
\nonumber
\qqq
with the use of causality. \,Stripping the resulting 
identity of the integrals against arbitrary functions 
$\,\dot{h}_{ta}$, \,we obtain the identity 
\qq
\beta\big\langle O^a_t\,O^b_{t}\big\rangle_{\hspace{-0.04cm}0}\,-\,
\beta\big\langle O^a_t\,O^b_{t'}\big\rangle_{\hspace{-0.04cm}0}\ 
=\ \theta(t-t')\int_{t'}^t\hspace{-0.1cm}\big\langle O^a_t\,R^b_{t''}
\big\rangle_{\hspace{-0.04cm}0}\,dt''
\ +\ \theta(t'-t)\int_{t}^{t'}\hspace{-0.1cm}\big\langle O^a_{t'}\,R^b_{t''}
\big\rangle_{\hspace{-0.04cm}0}\,dt''
\nonumber
\qqq
which is the integrated version of the differential relation between 
the dynamical 2-point correlation function and the response function:
\vskip 0.7cm

\noindent{\bf Proposition 5 \ (Fluctuation-dissipation theorem)}. \ 
For $\,t>t'$, 
\vskip -0.38cm
\qq
-\partial_t\,\big\langle O^a_t\,O^b_{t'}\big\rangle_{\hspace{-0.04cm}0}
\ =\ \beta^{-1}\big\langle O^a_t\,R^b_{t'}
\big\rangle_{\hspace{-0.04cm}0}\,.
\label{FDT}
\qqq
\vskip 0.6cm

\noindent Note the explicit factor $\,\beta\,$ in this identity.
Relations between the dynamical correlation functions and 
the response functions were used in recent years to extend the concept
of temperature to nonequilibrium systems \cite{CKP,Cilib}.
\vskip 0.3cm

\nsection{One-dimensional Langevin equation with flux solution}
\label{sec:1D}

Let us consider, as an illustration, the one-dimensional
Langevin equation of the form
\qq
\dot{x}\ =\ -\partial_xH_t(x)\,+\,\zeta_t
\label{LangD1}
\qqq
with $\,\big\langle\zeta_t\,\zeta_{t'}
\big\rangle=2\beta^{-1}\delta(t-t')\,$ (any force is a gradient
in one dimension). As before, $\,\sfx_t\,$
will represent the Markov process solving the SDE (\ref{LangD1}).
First, let us consider the time-independent case with a polynomial 
Hamiltonian $\,H(x)=ax^k+\dots\,$ with $\,a\not=0\,$ and the dots 
representing lower order terms.   

\noindent\begin{itemize}
\item If $\,k=0\,$ then, up to a linear change of variables, 
$\,\sfx_t\,$ is a Brownian motion and does not have an invariant 
probability measure.
\item If $\,k=1\,$ then $\,\sfx_t+at\,$ is, up 
to a linear change of variables, a Brownian motion and $\,\sfx_t\,$
still does not have an invariant probability measure.
\item If $\,k\geq2\,$ and is even then for $\,a>0\,$ 
the Gibbs measure $\,\mu_0(dx)=Z^{-1}\ee^{-\beta H(x)}dx\,$ provides 
the unique invariant probability measure of the process $\,\sfx_t$.
It satisfies the detailed balance condition $\,j(x)=0$, \,where
$\,j(x)\,$ is the probability current defined by Eq.\,(\ref{prcur}).
If $\,a<0$, however, then the Gibbs density $\,\ee^{-\beta H(x)}\,$ is
not normalizable\footnote{This leads to the breaking of the quantum-mechanical 
supersymmetry underlying the Fokker-Planck formulation of\\
\hspace*{0.6cm}the Langevin 
dynamics \cite{Witt,PS,TTNK}.}. In this case, the process $\,\sfx_t\,$
escapes to $\,\pm\infty\,$ in finite time with probability one and it has
no invariant probability measure.
\item If $\,k\geq3\,$ and is odd then the Gibbs density 
$\,\ee^{-\beta H(x)}\,$ is not normalizable. The process
$\,\sfx_t\,$ escapes in finite time to $\,-\infty\,$ if $\,a>0\,$
and to $\,+\infty\,$ if $\,a<0$, \,but it has a realization  
with the trajectories that reappear immediately from $\,\pm\infty$. 
\,Such a resuscitating process has a unique invariant probability measure
\qq
\mu_0(dx)\ =\ \pm N^{-1}\Big(\ee^{-\beta H(x)}\int\limits_{\mp\infty}^x
\ee^{\beta H(y)}dy\Big)dx\ \equiv\ \ee^{-\varphi_0(x)}dx
\label{fluxinv}
\qqq
with the density $\,\ee^{-\varphi_0(x)}=\CO(x^{-k+1})\,$ 
when $\,x\to\pm\infty\,$ and $\,N\,$ the (positive) 
normalization constant. The measure $\mu_0\,$ corresponds to a constant 
probability current $\,j(x)=\pm(\beta N)^{-1}\,$ and the model provides 
the simplest example on a nonequilibrium steady state with a constant 
flux.
\end{itemize}

\noindent Let us look closer at the last case. Adding the time-dependence 
and taking $\,\varphi_t\,$ as in Eq.\,(\ref{fluxinv}) but with $\,H_t\,$ 
replacing $\,H$, \,we obtain the Hatano-Sasa version of the Jarzynski 
equality (\ref{Jarz}) with $\,\CW=\CW^{ex}\,$ given by Eq.\,(\ref{HSW}). 
Suppose, in particular, that the time dependence of $\,H_t\,$ has the form 
(\ref{Ht}) with functions $\,O^a\,$ having compact support. Let us 
introduced also the deformed observables
\qq
\widehat{O^a}(x)\ =\ \frac{\int\limits_{\mp\infty}^xO^a(y)\,
\ee^{\beta H(y)}\,dy}{\int\limits_{\mp\infty}^x\ee^{\beta H(y)}\,dy}\,.
\nonumber
\qqq
Expanding the Jarzynski identity (\ref{Jarz}) to the second order 
in $\,h_{at}\,$ as in Sect.\,\ref{sec:FDT}, one obtains:
\vskip 0.7cm

\noindent{\bf Proposition 6 \ (Deformed fluctuation-dissipation relation).}
\ \ For $\,t>t'$,
\qq
-\partial_t\,\big\langle A^a_t\,O^b_{t'}\big\rangle_{\hspace{-0.04cm}0}
\ =\ \beta^{-1}\big\langle A^a_t\,R^b_{t'}\big\rangle_{\hspace{-0.04cm}0}\ 
\mp(\beta N)^{-1}\int(\partial_x O^b)(x)\,dx\,P_{t-t'}(x,dy)
\,A^a(y)\,,
\label{DFDT}
\qqq
where $\,P_t(x,dy)\,$ is the transition probability in the
stationary process and $\,A^a=O^a-\widehat{O^a}$.
\vskip 0.6cm

\noindent{\bf Remark 3.}\ \ It is easy to show directly, that 
Eq.\,(\ref{DFDT}) still holds if $\,A^a\,$ is replaced 
by $\,O^a$. \,Note that the term on the right hand side 
of (\ref{DFDT}) violating the standard fluctuation-dissipation 
theorem (\ref{FDT}) contains the constant flux of the probability 
current $\,j(x)\,$ as a factor. Proof of Proposition 6 
and of its version with $\,A^a\,$ replaced by $\,O^a\,$ 
will be given in \cite{ChFalkGaw}.
\vskip 0.6cm

The Langevin equation (\ref{LangD1}) with the flux solution arises
when one studies the tangent process for particles with inertia
moving in the one-dimensional homogeneous Kraichnan ensemble of 
velocities $\,v_t(y)\,$ with the covariance
\qq
\big\langle v_t(y)\,v_s(y')\big\rangle\ =\ \delta(t-s)\,D(y-y')\,,
\nonumber
\qqq
see Example 2. The position $\,y\,$ and the velocity $\,w\,$ of
such particles satisfy the SDE \cite{Bec}
\qq
\dot{y}\ =\ w\,,\qquad \dot{w}\ =\ \frac{_1}{^\tau}(-w+v_t(y))\,,
\nonumber
\qqq
where $\,\tau\,$ is the so called Stokes time measuring the time-delay
of particles with inertia as compared to the Lagrangian particles
that follow the flow. The separation between two infinitesimally
close trajectories of particles satisfies the equations 
\cite{WilkMehl}
\qq
\frac{_d}{^{dt}}\delta y\ =\ \delta w\,,\qquad\frac{_d}{^{dt}}\delta w\ 
=\ \frac{_1}{^\tau}(-\delta w+(\partial_y v_t)(y)\,\delta y)
\label{inert}
\qqq
and, similarly as in Example 5, we may replace $\,
\frac{_1}{^\tau}\partial_y v_t(y)\,$
on the right hand side by a white noise $\,\zeta(t)\,$ with the covariance
\qq
\big\langle\zeta_t\,\zeta_s\,\big\rangle\ =\ -\delta(t-s)\,
\tau^{-2}D''(0)\,,
\nonumber
\qqq
where the primes denote the spatial derivatives.
The ratio $\,x=\frac{\delta w}{\delta y}\,$ satisfies then the SDE
\qq
\dot{x}\ =\ -x^2\,-\,\frac{_1}{^\tau}x\,+\,\zeta_t
\label{SDEX}
\qqq
which has the form (\ref{LangD1}) with $\,H(x)=\frac{1}{3}x^3
+\frac{1}{2\tau}x^2$, \,a third order polynomial. The solution with 
the trajectories appearing at $\,+\infty\,$ after disappearing at 
$\,-\infty\,$ corresponds to the solution for $\,(\delta y,\delta w)\,$ 
with $\,\delta y\,$ passing through zero with positive speed. 
The top Langevin exponent for the random dynamical system (\ref{inert}) 
is obtained as the mean value of $\,x\,$ (which is the temporal 
logarithmic derivative of $\,|\delta y|$) in the invariant probability 
measure (\ref{fluxinv}) with constant flux \cite{WilkMehl}.
\vskip 0.2cm

A very similar SDE arose earlier \cite{Halp} in the one-dimensional 
Anderson localization in white-noise potential $\,V(y)\,$ where one 
studies the stationary Schr\"odinger equation 
\qq
-\psi''(y)\,+\,V(y)\,\psi(y)\ =\ E\,\psi(y)\,.
\nonumber
\qqq
By setting $\,x=\psi'/\psi$, 
\,one obtains then the evolution SDE
\qq
x'\ =\ -x^2\,-\,E\,+\,V(y)\,.
\label{SDEY}
\qqq
that has an invariant probability measure with constant flux, as already
noticed in \cite{Halp}. The expectation value of $\,x\,$ in that measure
may be expressed by the Airy functions \cite{LGP}. It
gives the (top) Lyapunov exponent which is always positive,
reflecting the permanent localization in one dimension.
The SDE (\ref{SDEX}) may be obtained from (\ref{SDEY})
but taking in the latter $\,E=-\frac{1}{4\tau^2}\,$ and by the substitutions 
$\,x-\frac{t}{2\tau}\mapsto x$, $\,V\mapsto\zeta\,$ and $\,y\mapsto t$. 
\,This shifts the Lyapunov exponent down by $\,-\frac{1}{2\tau}\,$ and the 
top exponent for the inertial particles may have both signs \cite{WilkMehl}.
\vskip 0.3cm

\nsection{Detailed fluctuation relation}
\label{sec:DFR}

For a general pair of forward and backward diffusion processes 
(\ref{E0}) and (\ref{inversion}), it is still possible 
to obtain identities resembling the generalized detailed balance 
relation (\ref{DB1}) at the price of adding constraints on the 
process trajectories. Let us introduce a functional  $\,\CW'\,$ 
of the backward process by mimicking the definition 
(\ref{W0T}) of $\,\CW\,$ for the forward process:
\qq
\CW'\,=\,\Delta\varphi'\,+
\int\limits_0^T\CJ'_{t}\,dt
\nonumber
\qqq
with
\qq
\CJ'_{t}\,=\,2\,\hat u'_{t,+}
(\sfx'_{t})\cdot d'^{-1}_{t}
(\sfx'_{t})\,\dot{\sfx}'_{t}\,-\,2\,\hat u'_{t,+}
(\sfx'_{t})\cdot d'^{-1}_{t}(\sfx'_{t})
\,u'_{t,-}(\sfx'_{t})\,-\,(\nabla\cdot u'_{t,-})(\sfx'_{t})\,,
\nonumber
\qqq
see Eq.\,(\ref{js}). Since the time inversion is involutive, the mirror 
version of the identity (\ref{BI0}), 
\qq
\mu'_0(dx')\,\,\bE'^{0,T}_{x',y'}\,\,\CF'(\sfx')\,\,
\ee^{-\CW'(\sfx')}\,\,dy'\ =\
\mu_0(dy'^*)\,\,\bE^{0,T}_{y'^*\hspace{-0.1cm},x'^*}
\,\,\tilde{\CF'}(\sfx)\,\,dx'^*\hspace{0.02cm}
dx'^*\,,
\label{BI0'}
\qqq
must also hold. Taking $\,x'=y^*$, $\,y'=x^*\,$ and 
$\,\CF'=\tilde\CF\,\ee^{-\tilde\CW}$, 
\,we infer that the compatibility of identities (\ref{BI0}) and (\ref{BI0'}) 
imposes the equality
\qq
\CW'\ =\ -\tilde\CW\,,
\label{sym}
\qqq
which may be also checked directly. \,We infer that, whatever the time 
inversion used in their definition, the entropy-production functionals 
$\,\CW\,$ for the forward and the backward processes are related by 
the natural time inversion. The replacement in Eq.\,(\ref{BI0}) of 
the functional $\,\CF(\sfx)\,$ by the functional $\,\CF(\sfx)\,
\delta(\CW(\sfx)-W)\,$ including the constraint fixing the value of 
$\,\CW$, \,leads then to
\vskip 0.8cm

\noindent{\bf Proposition 7 \ (Detailed fluctuation relation).}
\vskip -0.1cm

\qq
\mu_0(dx)\,\,\bE_{x,y}^{0,T}\,\,\CF(\sfx)\,\,
\delta\big(\CW(\sfx)-W\big)\,\,dy\ 
=\ \mu'_0(dy^*)\,\,\ee^{\,\CW}\,\,
\bE'^{0,T}_{y^*\hspace{-0.1cm},x^*}
\,\,\tilde\CF(\sfx')\,\,
\delta\big(\CW'(\sfx')+W\big)\,\,dx^*\,.
\label{DFR}
\qqq
\vskip 0.2cm

\noindent The primes on the right hand side may be dropped in the 
time-reversible case if, additionally, $\,\varphi_0=\varphi'_0\,$
and $\,\varphi_T=\varphi'_T$.
\vskip 0.2cm

\hspace{10cm}$\Box$
\vskip 0.6cm

\noindent A relation of this type, named the ''detailed fluctuation 
theorem'', was established in \cite{Jarz4} in a setup of the Hamiltonian 
dynamics. It is close in spirit to the earlier observation made
for the long-time asymptotics of deterministic dynamical systems
in \cite{Gall1}. We shall view Proposition 7 as a source of fluctuation 
relations that hold for the diffusion processes (\ref{E0}), 
including the Jarzynski equality (\ref{Jarz}) already discussed and
various identities that appeared in the literature in different
contexts, see \cite{Jarz4,Crooks1,Crooks2,Kurchan1}. Taking, in 
particular, $\,\CF\equiv1\,$ in Eq.\,(\ref{DFR}) and introducing the
joint probability distributions of the end-point of the process
and of the entropy production functional $\,\CW$,
\qq
\bE_{x,y}^{0,T}\,\delta(\CW(\sfx)-W)\,dy\,dW\,=\,P_{0,T}(x,dy,dW)\,,
\qquad
\bE'^{0,T}_{x,y}\,\delta(\CW'(\sfx')-W')\,=\,P'_{0,T}(x,dy,dW')\,,
\nonumber
\qqq
we obtain 
\vskip 0.7cm

\noindent{\bf Corollary 6.}
\qq
\mu_0(dx)\,\,P_{0,T}(x,dy,dW)\ =\ \mu_T(dy)\,\,
\ee^{\,W}\,\,P'_{0,T}(y^*,dx^*,d(-W))\,.
\label{cor6}
\qqq
\vskip 0.2cm

\noindent This may be viewed as an extension to a general diffusive 
SDE (\ref{E0}) of the detailed balance relation (\ref{DBP}), or of its 
generalization (\ref{DB1}). In particular, when the 
backward process is obtained by the complete reversal of 
Sect.\,(\ref{sec:comprev}) with $\,\CW\equiv0$, the latter relation
reduces to Eq.\,(\ref{DB1}) with both sides multiplied by 
$\,\delta(W)dW$. 
\vskip 0.5cm

In the case when the measures $\,\mu_0\,$
and $\,\mu'_0\,$ are normalized, \,Proposition 7 gives rise, upon 
integration over $\,x\,$ and $\,y$, \,to 
a detailed fluctuation relation between the forward and the backward
processes with the initial points sampled with measures $\,\mu_0\,$
and $\,\mu'_0$, \,respectively:
\vskip 0.7cm

\noindent{\bf Corollary 7.}
\qq
\Big\langle\,\CF\,\,\delta\big(\CW-W\big)
\Big\rangle\ 
=\ \ee^W\,\Big\langle\,\tilde\CF\,\,
\delta\big(\CW'+W\big)\,\Big\rangle^{\hspace{-0.06cm}\prime}\,. 
\nonumber
\qqq
\vskip 0.6cm

\noindent Finally, taking $\,\CF=1\,$ in the latter identity and denoting
\qq
p_{0,T}(dW)\,=\,\Big\langle\,\delta\big(\CW-W\big)\,
\Big\rangle\,dW\,,\qquad
p'_{0,T}(dW')\,=\,\Big\langle\,\delta\big(\CW'-W'\big)\,
\Big\rangle^{\hspace{-0.06cm}\prime}\,dW'\,,
\nonumber
\qqq
we obtain
\vskip 0.7cm

\noindent{\bf Corollary 8 \ (Crooks relation)} \cite{Crooks1,Crooks2}.
\qq
p_{0,T}(dW)\ =\ \ee^{\,W}\,\,p'_{0,T}(d(-W))\,.
\label{Crooks}
\qqq
\vskip 0.6cm

\noindent Note that $\,p_{0,T}(dW)\,$ is the distribution  
of the random variable $\,\CW\,$ if the time-zero values 
of the forward process $\,\sfx_t\,$ are distributed with the measure 
$\,\mu_0\,$ and, similarly, $\,p'_{0,T}(dW')\,$ is the 
distribution of the random variable $\,\CW'\,$ if 
$\,\sfx'_0\,$ is distributed with the measure $\,\mu'_0$. 
\,In particular, in the time-reversible case, 
$\,p'_{0,T}(dW)=p_{0,T}(dW)\,$ if $\,\varphi'_0=\varphi_0\,$ and 
$\,\varphi'_T=\varphi_T$. \,Finally, note that integrating the Crooks 
relation (\ref{Crooks}) multiplied by $\,\ee^{-W}\,$ over $\,W$, \,one 
recovers the Jarzynski equality (\ref{Jarz}).
\vskip 0.3cm

\nsection{Special cases}
\label{sec:spcases}
\subsection{Deterministic case}

As already explained in Sect.\,\ref{sec:u+=0} and 13, taking 
$\,\hat u_{t,+}=0\,$ and 
$\,u_{t,-}=\hat u_t\,$ leads in the limit of the deterministic 
dynamics (\ref{E0det}) to the expression (\ref{df}) for $\,\CW$.
\,The time-reversed dynamics corresponds to the  
vector fields of Eqs.\,(\ref{hatback}). It reduces in the 
deterministic case to the ODE (\ref{detback}). The functional
$\,\CW'\,$ of the backward process, that could be also found 
from the relation (\ref{sym}), takes the form
\qq
\CW'\,=\,\Delta\varphi'\,+\,
\int\limits_0^T\big[(\nabla\ln\sigma)(\sfx'_{t})
\cdot(\dot{x}^\prime_{t}-u'_{t,-})\,-
\,(\nabla\cdot u'_{t,-})(\sfx'_{t})\big]dt\,.
\nonumber
\qqq
In the deterministic limit, this simplifies to the expression 
\qq
\CW'\,=\,\Delta\varphi'\,-\,
\int\limits_0^T(\nabla\cdot u'_{t})(\sfx'_{t})\,dt
\nonumber
\qqq
which is of the same form as Eq.\,(\ref{df}) for $\,\CW$. 
\,Proposition 7 and Corollaries 6,7 and 8 still hold in the deterministic 
limit. In particular, in the time-reversible deterministic case with 
$\,u'=u\,$ and $\,\varphi_T=\varphi_0=\varphi'_0$, \,the fluctuation 
relation (\ref{Crooks}) reduces to
\vskip 0.6cm

\noindent{\bf Corollary 9 \ (Evans-Searles transient fluctuation theorem)} 
\cite{EvSear,EvSear2}
\qq
p_{0,T}(dW)\ =\ \ee^W\,\,p_{0,T}(d(-W))\,.
\label{EvSear}
\nonumber
\qqq
\vskip 0.2cm

\noindent The latter relation may also be proven directly by a change 
of the integration variables $\,\sfx_0\mapsto\sfx_t\,$ \cite{EvSear2}.

\subsection{Reversed protocol case}

For the reversed protocol time inversion of Sect.\,\ref{sec:revprot} 
and Example 11 that corresponds to the choice (\ref{revprot}),
the backward process is given by Eq.\,(\ref{revprot'}) and 
\qq
\CW'\ =\ \Delta\varphi'\,+\,
2\int\limits_0^T\hat u'_{t}(\sfx'_{t})\cdot d'^{-1}_{t}(\sfx'_{t})\,
\dot{\sfx}'_{t}\,dt
\nonumber
\qqq
and has the same form as $\,\CW$, \,see Eq.\,(\ref{revprotW}).
\,For such a time inversion with $\,x^*\equiv x$, \,employed already 
in the stationary context in \cite{LebowSp}, the fluctuation 
relation (\ref{Crooks}) for the choice of $\,\varphi_t\,$ such 
that $\,L_t^\dagger\ee^{-\varphi_t}=0\,$ was established in 
\cite{ChChJarz}.

\subsection{Current reversal case} 

For the time inversion (\ref{CRsplit}) discussed in Sect.\,\ref{sec:currev}
and Example 12, the functional $\,\CW'\,$ of the backward process is given 
by the expression of the same form as Eq.\,(\ref{HSW}):
\qq
\CW'\ =\ \int\limits_0^T(\partial_{t}
\varphi'_{t})(\sfx'_{t})\,dt
\nonumber
\qqq
for $\,\varphi'_{t}(x)=(\varphi_{t^*}+\ln{\sigma})(x^*)$.
\,The fluctuation relation (\ref{Crooks}) for this type
of time inversion (with $\,x^*\equiv x$) \,was proven by 
in \cite{ChChJarz}. Integrated against $\,\ee^{-W}$, 
\,Eq.\,(\ref{Crooks}) reduces to the Hatano-Sasa case of the Jarzynski 
equality (\ref{Jarz}) that we discussed in Example 12.

\subsection{Langevin dynamics case}

Recall that for the Langevin dynamics (\ref{Lang}), the backward process
obtained by using a canonical time inversion defined by Eqs.\,(\ref{cano0}) 
and (\ref{cano}) is also of the Langevin type with
\qq
u'_t\,=\,-\Gamma\nabla H'_t\,+\,\Pi\nabla H'_t\,+\,G'_t\,,\qquad
\label{u'lang}
\qqq
where $\,H'_t(x)=H_{t^*}(rx)$, $\,G'_t(x)=-rG_{t^*}(rx)$. \,The white
noise $\,\zeta'_t=\pm r\zeta_{t^*}\,$ has the same distribution as $\,\zeta_t$.
\,Consequently, for $\,\varphi'_t=\beta(H'_t-F'_t)$, \,the functional
$\,\CW'\,$ is given by the primed version of Eq.\,(\ref{JarzW0})
and is equal to the dissipative work (in the $\beta^{-1}$ units).
\vskip 0.2cm

If, instead of the canonical time inversion, we use the reversed 
protocol with $\,x^*\equiv x\,$ then the backward process is again 
the Langevin dynamics with $\,u'_t\,$ given by Eq.\,(\ref{u'lang}), 
except that this time $\,H'_t(x)=H_{t^*}(x)\,$ and $\,G'_t(x)
=G_{t^*}(x)$. \,The white noise $\,\zeta'_t=\zeta_{t^*}\,$
has again the same distribution as $\,\zeta_t$. \,The functional 
$\,\CW'\,$ is given in that case by the primed version
of the Eq.\,(\ref{WtotL}). The two time inversions lead to the 
equivalent backward processes for the Langevin-Kramers equation but, 
as already mentioned, $\,\CW^{tot}\,$ is not well defined in the case
of the reversed protocol.
\vskip 0.2cm

Finally, if we apply the current-reversal time inversion 
(\ref{CRsplit}) with $\,x^*\equiv x\,$ to the Langevin dynamics 
(\ref{Lang}) with $\,G_t\equiv0\,$ by setting 
$\,\varphi_t=\beta(H_t-F_t)=\varphi'_{t^*}=\beta(H'_{t^*}
-F'_{t^*})\,$ for $\,H'_t(x)=H_{t^*}(x)$,
\,the drift of the backward dynamics becomes
\qq
u'_{t}\ =\ -\,\Gamma\nabla H'_{t}\,
-\,\Pi\nabla H'_{t}
\nonumber
\qqq
and has the changed sign of the antisymmetric matrix $\,\Pi\,$ 
with respect to the forward process. The white noise $\,\zeta'_t
=\pm\zeta_{t^*}$. \,Here both $\,\CW\,$ and $\,\CW'\,$ have the form 
of the dissipative work. 
\vskip 0.3cm

\nsection{Transient versus stationary fluctuation relations}
\label{sec:stationFR}

The fluctuation relations considered up to now dealt with the
quantities related to finite-time evolution in a random process
that, in general, was not stationary. Such simple relations, whose 
prototypes where the Evans-Searles fluctuation relation \cite{EvSear} 
or the Jarzynski equality \cite{Jarz1} are called {\bf transient} 
fluctuation relations. On the other hand, as was recalled in 
Introduction, Gallavotti and Cohen have established in \cite{GC} 
a fluctuation relation for quantities pertaining to the long-time 
evolution in stationary deterministic dynamical systems of chaotic type 
and similar relations were subsequently obtained for the Langevin 
dynamics and Markov processes in \cite{Kurchan1} and \cite{LebowSp}. 
Such fluctuation relations, that are commonly termed
{\bf stationary}, are usually more difficult to establish than
the transient ones and require some non-trivial work that involves 
the existence and the properties of the stationary regime of the dynamics. 
Such properties are in general harder to establish in the non-random case 
than in the random one. Also, in the random case, the invariant measure
of the process, if exist, is usually smooth. It could be
used as the measure $\,\mu_0(dx)=\ee^{-\varphi_0(x)}dx=\mu_T(dx)=
\mu'_0(dx^*)\,$ in the definition (\ref{W0T}), leading to the exact detailed 
fluctuation relation (\ref{DFR}) pertaining to the stationary
evolution. On the other hand, in the dissipative deterministic
systems, the invariant (SRB) measures are not smooth, so that
they may not be used this way and the exact stationary fluctuation
relations may be obtained only in the asymptotic long-time regime.  
Let us discuss briefly a formal relation between such asymptotic
fluctuation relations and the transient ones, sweeping under the rug 
the hard points.
\vskip 0.3cm

We shall consider the stationary case of the SDE (\ref{E0}), with
$\,u_t\equiv u\,$ and $\,D_t(x,y)\equiv D(x,y)$. \,Under precise
conditions, the Markov process $\,\sfx_t\,$ that has decaying dynamical 
correlations and attains at long times the steady state independent 
of the initial (or/and final) position \cite{Hashm,Kunita}. In such 
a situation, the distribution of the functional $\,\CW\,$ is expected 
(and may often be proven with some work) to take for long time $\,T\,$ 
and for $\,\CW/T=\CO(1)\,$ the large deviation form 
\qq
P_{0,T}(x,dy,dW)\ \ \propto\ \ \ee^{-T\,\zeta(\CW/T)}\,dy\,dW
\label{ld}
\qqq
independent of $\,x\,$ and $\,y$. \,The function $\,\zeta\,$ is called
the large deviations rate function. It is convex and has vanishing minimum.
More exactly, the relation (\ref{ld}) means that
\qq
&\displaystyle{-\sup\limits_{w\in\CI}\zeta(w)\ \leq\ 
\mathop{\underline{\lim}}\limits_{T\to\infty}\frac{1}{T}\ln
\int\limits_{T\CI}P_{0,T}(x,y|\CW)\,d\CW}&\cr
&\displaystyle{\leq\ 
\mathop{\overline{\lim}}\limits_{T\to\infty}\frac{1}{T}\ln
\int\limits_{T\CI}P_{0,T}(x,y|\CW)\,d\CW\ \leq\ 
-\inf\limits_{w\in\CI}\zeta(w)}&
\nonumber
\qqq
for any interval $\,\CI\,$ in the real line. In particular,
in the limit $\,T\to\infty$, \,the distribution of $\,\CW/T\,$ 
concentrates at the non-random value $\,w_{\hspace{-0.01cm}0}\,$ 
where the rate function $\,\zeta\,$ attains its minimum. 
With similar assumptions about the inverse process, we shall denote by 
$\,\zeta'\,$ the large deviation rate function of the functional
$\,\CW'$. \,The detailed fluctuation relation (\ref{cor6}) implies 
then immediately, if the boundary term $\ \varphi'_0(y^*)/T
=(\varphi_T(y)+\ln{\sigma(y)})/T\ $ converges to zero when
$\,{T\to\infty}0$, \,a relation between the rate functions 
$\,\zeta\,$ and $\,\zeta'$:
\vskip 0.8cm

\noindent{\bf Corollary 3 \ (Stationary fluctuation relation).}
\vskip -0.38cm
\qq
\zeta(w)\ =\ \zeta'(-w)\,-\,w
\label{sfr}
\qqq  
\vskip 0.7cm

\noindent Eq.\,(\ref{sfr}) connects the statistics of large deviations 
of $\,\CW\,$ for the forward and for the backward stationary stochastic 
processes. Note that the equality $\,\zeta'\geq0\,$ implies that 
the asymptotic value $\,w_{\hspace{-0.01cm}0}\,$ of $\,\CW/T\,$ is 
non-negative. This conclusion may be also drawn from the $2^{\rm nd}$ law 
(\ref{2law}). \,In the special case of a stationary time-reversible 
dynamics, the inverse process coincides with the direct one so that 
$\,\zeta'=\zeta$. \,Eq.\,(\ref{sfr}) compares then the large deviations 
of $\CW/T\,$ of opposite signs in the forward process. In particular, 
it states that the probability that $\,\CW/T\,$ takes values opposite 
to the most probable ones around $\,w_{\hspace{-0.01cm}0}\,$ is 
suppressed by the exponential factor $\,\ee^{-T\,w_{\hspace{-0.02cm}_0}}\,$
for large times $\,T$. 
\vskip 0.2cm

Recall from the definition (\ref{W0T}) that $\,\CW\,$ differs
from the extensive quantity $\,\int\limits_0^T\CJ_t\,dt\,$
by a boundary term which should not contribute to the large
deviations if $\,\varphi_T\,$ stays bounded, although presence
of such terms may change the time-scales on which the large 
deviation regime is effectively visible. On the contrary, unbounded 
$\,\varphi_T\,$ may give contributions to the large deviations 
statistics \cite{vZonC,BGallGZ,WEvSear,PRV}. For the deterministic
dynamics where $\,\int\limits_0^T\CJ_t\,dt=-\int\limits_0^T
(\nabla\cdot u)(\sfx_t)\,dt\,$ is the phase-space contraction along 
the trajectory, see Eq.\,(\ref{CJdet}), the identity (\ref{sfr}) with 
$\,\zeta'=\zeta\,$ is essentially the original Gallavotti-Cohen 
fluctuation relation \cite{GC,Gall1} established rigorously 
by the authors for the reversible Anosov dynamical 
systems with discrete time. For such systems, the thermodynamic 
formalism \cite{Ruelle2,GBG} may be used to prove the existence 
of the stationary (SRB) measure and of the large deviations regime for 
the phase-space contraction, see also \cite{Ruelle1} for a somewhat 
different approach. In \cite{Kurchan}, the fluctuation relation 
(\ref{sfr}) was discussed for the Langevin-Kramers dynamics, see also 
\cite{LebowSp,Kurchan1}. Its version considered here for a general
stationary diffusion process is equivalent in the case of vanishing 
time-inversion-odd drift $\,u_-\,$ to the fluctuation relation
discussed in \cite{LebowSp}, see Eq.\,(5.8) therein.
\vskip 0.3cm

As another (although related) example of how the transient fluctuation
relations yield stationary ones involving large deviations, let us recall 
the case of the tangent process in the homogeneous Kraichnan model 
leading to the It\^o multiplicative SDE (\ref{tprI}) (or the Stratonovich 
SDE (\ref{E0W}) equivalent to it) and defining the matrix-valued 
process $\,{\sfX}_t$. \,We have established for it the transient 
fluctuation relation (\ref{KrFR}) that may be rewritten as the identity
\qq
\bE^0_1\,\,\det(\sfX_T)\,f(\sfX_T)\ =\ \bE^0_1\,\,f(\sfX_T^{-1})\,.
\label{KrFR1}
\qqq                  
for functions $\,f\,$ of real $\,d\times d\,$ matrices with positive
determinant. Such matrices $\,X\,$ may be cast into
the form 
\qq
X\ =\ O'\,{\rm diag}(\ee^{\rho_1},\dots,\ee^{\rho_d})\,O^{-1}
\label{strexp}
\qqq
with a diagonal matrix of non-increasing positive entries sandwiched 
between two orthogonal ones. Note that $\,\ln\det X=\sum\rho_i$. \,The so 
called {\bf stretching exponents} $\,\rho_1\geq\,\cdots\,\geq\rho_d\,$
are uniquely defined by Eq.\,(\ref{strexp}). Consider functions 
$\,f(X)\,$ that are left- and right-invariant under the action 
of the orthogonal group $\,O(d)$. \,They may be viewed as functions of 
the vector $\,\vec{\rho}\,$ of the stretching exponents.  The distribution 
$\,P_T(d\vec{\rho})\,$ of such exponents is defined 
by the relation
\qq
\bE^0_1\,f(\sfX_T)\ =\ \int\limits_{\rho_1\geq..\geq\rho_d}\hspace{-0.2cm} 
f(\vec{\rho})\,P_T(d\vec{\rho})\,.
\nonumber
\qqq
The identity (\ref{KrFR1}) implies then that
\qq
P_T(d\vec{\rho})\,\,\ee^{\sum\rho_i}\ =\ P_T(d(-\antivec{\rho}))\,,
\label{KrFR2}
\qqq
where $\,-\antivec{\rho}=(-\rho_d,\dots,-\rho_1)\,$ is the vector
of the stretching exponents of the matrix $\,X^{-1}$. \,In few 
particular situations (e.g. in the isotropic case), it has
been established that for long times and $\,\vec{\rho}/T=\CO(1)$, 
\,the distribution of the stretching exponents takes the large deviation 
form
\qq
P_T(d\vec{\rho})\ \ \propto\ \ \ee^{-T\,Z(\vec{\rho}/T)}d\vec{\rho}
\nonumber
\qqq 
and the identity (\ref{KrFR2}) implies then the stationary
fluctuation relation
\qq
Z(\vec{\sigma})\,-\,\sum\limits_{i=1}^d\sigma_i\ =\ Z(-\antivec{\sigma})\,,
\label{MGC}
\qqq
see \cite{aniso}. Since $\,-\sum\rho_i\,$ represents the phase-space
contraction $\,-\ln\det\sfX_t\,$ in the Kraichnan model, the relation
(\ref{MGC}) may be viewed as a modified Gallavotti-Cohen identity 
(\ref{sfr}) for the homogeneous Kraichnan model. The modification
goes in two directions. On one hand side, the original Gallavotti-Cohen 
relation involved the deterministic dynamics, whereas the relation 
(\ref{MGC}) pertains to random Kraichnan dynamics. On the other hand, it
refers to the ``multiplicative'' large deviations for the vector
$\,\vec{\rho}\,$ of the stretching exponents containing more detailed 
information than the phase-space contraction represented by $\,-\sum\rho_i$. 
\,For example, the most probable values of the {\,\bf stretching rates} 
$\,\sigma_i=\rho_i/T\,$ for which $\,Z(\vec{\sigma})=0\,$ define the 
{\bf Lyapunov exponents} $\,\lambda_i\,$ whereas the most probable 
phase-space contraction rate is equal to the negative of their sum. 
We shall see in the next section how to extend such multiplicative 
fluctuation relations to the general diffusive processes. The source of such 
an extension resides in transient relations that may be proven for general 
random or deterministic dynamical systems by a simple change-of-variables 
argument {\it \`a la} Evans-Searles \cite{EvSear2}, as first indicated
in \cite{BFF}. 
\vskip 0.3cm

\nsection{Multiplicative fluctuation relations}
\label{sec:MFR}

As we have mentioned above, the SDE (\ref{E0}) defining the diffusive
process $\,\sfx_t\,$ may be used to induce other diffusive processes, the 
simplest example being the tangent process $\,(\sfx_t,\sfX_t)\,$ 
introduced in Sect.\,\ref{sec:tanproc} and satisfying the SDEs
\qq
\dot{x}\ =\ u_t(x)\,+\,v_t(x)\,,\qquad
\dot{X}\ =\ U_t(x,X)\,+\,V_t(x,X)
\nonumber
\qqq
with
\qq
U^i_{t\ j}(x,X)\,=\,(\partial_k u^i_t)(x)\,X^k_{\,\ j}\,,
\qquad V^i_{t\ j}(x,X)\,=\,(\partial_k v^i_t)(x)\,X^k_{\,\ j}\,,
\nonumber
\qqq
see Eq.\,(\ref{tpr}). The covariance of the white noise vector field
$\,(v_t,V_t)\,$ is given by the relations (\ref{dep2}) and
\qq
&&\hbox to 3.6cm{$\langle\,v^i_t(x)\,V^k_{s\ l}(y,Y)\,\rangle$\hfill}
=\ \ \delta(t-s)\ 
\partial_{y^m}D^{ik}_t(x,y)\,Y^m_{\,\,\,\,l}\,,\cr\cr 
&&\hbox to 3.6cm{$\langle\,V^p_{t\ r}(x,X)\,v^j_s(y)\,\rangle$\hfill}
=\ \ \delta(t-s)\ 
\partial_{x^n}D^{pj}_t(x,y)\,X^n_{\,\ r}\,,\cr\cr
&&\hbox to 3.6cm{$\langle\,V^p_{t\ r}(x,X)\,V^k_{s\ l}(y,Y)\,
\rangle$\hfill}=\ \ \delta(t-s)\ \partial_{x^n}\partial_{y^m}D^{pk}_t(x,y)
\,X^n_{\,\ r}\,Y^m_{\,\,\,\,l}\,.
\nonumber
\qqq
One may now apply the theory developed
above for general diffusion processes to the case of tangent process.
As an example, let us consider the natural time inversion 
of Sect.\,\ref{sec:nat} corresponding to the trivial splitting
$$(u_{t,+},\,U_{t,+})\,=\,0\,,\qquad(u_{t,-},\,U_{t,-})\,=\,(u_{t},\,
U_{t})$$
and to the involution
$$(x,X)^*\,=\,(x^*,X^*)\qquad{\rm with}\qquad (X^*)^i_{\,\ j}\,=\,(\partial_k
{x^*}^i)(x)\,X^k_{\,\ j}\,.$$
The backward process
$\,(\sfx'_{t},\sfX'_{t})\,$ satisfies in this case the SDE
\qq
\dot{x'}\ =\ u'_{t}(x')\,+\,v'_{t}(x')\,,\qquad
\dot{X'}\ =\ U'_{t}(x',X')\,+\,V'_{t}(x',X')
\nonumber
\qqq
with 
\qq
&&u'^i_{t}(x)\,=\,-(\partial_k{x^*}^i)(x^*)\,u^k_{t^*}(x^*)\,,\cr\cr
&&U'_{t}(x,X)\,=\,-(\partial_k\partial_m{x^*}^i)(x^*)\,
(X^*)^m_{\,\ j}\,u^k_{t^*}(x^*)\,-\,(\partial_m{x^*}^i)(x^*)
\,U^m_{t^*\,j}(x^*,X^*)\cr\cr
&&\hspace{1.5cm}=\,(\partial_nu'^i_t)(x)\,X^n_{\,\ j}
\nonumber
\qqq
and, similarly,
\qq
&&v'^i_{t}(x)\,=\,\pm(\partial_k{x^*}^i)(x^*)\,v^k_{t^*}(x^*)\,,
\qquad V'_{t}(x,X)\,=\,(\partial_nv'^i_t)(x)\,X^n_{\,\ j}\,.\quad
\nonumber
\qqq
Note that the backward process $\,(\sfx'_{t},\sfX'_{t})\,$
defined this way coincides with the tangent process of $\,\sfx'_{t}$.
\,Eqs.\,(\ref{stand}) applied to the case at hand give:
\qq
\big(\hat u^i_{t,+}(x)\,,\ (\hat U_{t,+})^k_{\ l}(x,X)\big)\ =\ 
-\frac{_{d+1}}{^2}\big(\partial_{y^n}D^{in}(x,y)|_{y=x}\,,\ 
\partial_{x^n}\partial_{y^m}D^{km}(x,y)|_{y=x}\,X^n_{\,\ l}\big)
\cr\cr 
=\ -\frac{_{d+1}}{^2}\,\big(0\,,\ (X^{-1})^r_{\ p}\big)\,
\left(\begin{matrix}d^{ij}_t(x)&
\partial_{y^m}D^{ik}_t(x,y)|_{y=x}\,X^m_{\,\,\,\,l}\cr
\partial_{x^n}D^{pj}_t(x,y)|_{y=x}\,X^n_{\,\ r}&
\partial_{x^n}\partial_{y^m}D^{pk}_t(x,y)|_{y=x}\,X^n_{\,\ r}\,
X^m_{\,\,\,\,l}\end{matrix}\right)
\nonumber
\qqq
in the matrix notation, where the matrix on the right hand side is
the counterpart of $\,\big(d^{ij}_t(x)\big)\,$ for the tangent process. 
Substituting the above expression to the definition (\ref{js}),
we infer that
\qq
\CJ_t\ =\ -(d+1)(\sfX_t^{-1})^r_{\ p}\,\dot\sfX^p_{t\,\,r}
\ +\ (d+1)(\sfX_t^{-1})^r_{\ p}\,\partial_nu^p_t(\sfx_t)\,\sfX^n_{t\,\,r}
\ -\ (d+1)\,\partial_nu^n_t(\sfx_t)\,\,\cr
=\ -(d+1)\,\frac{d}{dt}\ln\det\sfX_t\,.
\nonumber
\qqq
The relation (\ref{P1P}) of Sect.\,\ref{sec:forwback} gives then 
for the case of the tangent process the identity
\qq
dx\,dX_0\,\,P_{0,T}(x,X_0;dy,dX)\ (\det X_0)^{-(d+1)}(\det X)^{d+1}\ 
=\ dy\,dX\,\,P'_{0,T}(y^*,X^*;dx^*,dX^*_0)
\nonumber
\qqq
that may be viewed as an extension of the relation 
(\ref{KrFR0}) obtained in Example 5 for the homogeneous Kraichnan 
process to a general diffusive process. Similarly as in Example 5,
we infer from the above equation the {\bf multiplicative fluctuation 
relation}
\qq
dx\,\,P_{0,T}(x,1;dy,dX)\,(\det X)\ =\ dy\,\,
P'_{0,T}(y^*,1^*;dx^*,d(X^{-1})^*)\,.
\label{MFR}
\qqq  
Suppose that we are given a Riemannian metric $\,\gamma\,$ on $\,\bR^d\,$
(for example the usual flat one). Since the matrix $\,X=\sfX_T\,$ maps the 
tangent space at $\,x=\sfx_0\,$ to the one at $\,y=\sfx_T$, 
\,see Eq.\,(\ref{tpr}), it is natural to define the stretching
exponents $\,\vec{\rho}\,$ of $\,X\,$ by the relation (\ref{strexp}) 
with $\,O\,$ and $\,O'\,$ mapping the canonical basis of $\,\bR^d\,$ 
into a basis orthonormal with respect to the metric $\,\gamma(x)\,$ 
and $\,\gamma(y)$, \,respectively. The joint probability distribution
$\,P_{0,T}(x,dy,d\vec{\rho})\,$ of the end-point of the process 
$\,\sfx_t\,$ and of the stretching exponents of $\,\sfX_t\,$ is then 
given by the relation
\qq
\int f(X)\,\,P_{0,T}(x,1,dy,dX)\ =\int\limits_{\rho_1\geq..\geq\rho_d}
\hspace{-0.2cm}f(\vec{\rho})\,\,P_{0,T}(x,dy,d\vec{\rho})
\nonumber
\qqq
for functions $\,f(X)\,$ left- and right-invariant under the action 
of the orthogonal groups preserving, respectively, the metric 
$\,\gamma(x)\,$ and $\,\gamma(y)$.
\,Similarly we introduce the kernels $\,P'_{0,T}(x',dy',d\vec{\rho}')\,$ 
using the transition probabilities of the backward process and the metric 
$\,\gamma'\,$ obtained from $\,\gamma\,$ by the involution $\,x\mapsto x^*$.
\,Eq.\,(\ref{MFR}) implies then the identity
\qq
v_\gamma(dx)\,\,P_{0,T}(x,dy;d\vec{\rho})\,\,\ee^{\,\sum\limits_i\rho_i}\ 
=\ v_\gamma(dy)\,\,P'_{0,T}(y,dx^*;d(-\antivec{\rho}))\,.
\nonumber
\qqq
where $\,v_\gamma(dx)\,$ is the metric volume measure.
For the stationary dynamics, we may expect the emergence of
the large deviations regime for the stretching rates with
\qq
P_{0,T}(x,dy;d\vec{\rho})\ \ \cong\ \ \ee^{-T\,Z(\vec{\rho}/T)}\,dy
\,d\vec{\rho}
\nonumber
\qqq
for large $\,T\,$ and $\,\vec{\rho}/T=\CO(1)$, \,and similarly 
for the backward process. One obtains then the identity
\qq
Z(\vec{\sigma})\ -\ \sum\limits_{i=1}^d\sigma_i\ =\ Z'(-\antivec{\sigma})\,.
\label{MGCgen}
\qqq
As usually, the rate function $\,Z'\,$ for the backward process may 
be replaced by $\,Z\,$ for a time-reversible dynamics. The relation 
(\ref{MGCgen}) generalizes the fluctuation relation (\ref{MGC}) obtained 
for the Lagrangian flow in the homogeneous Kraichnan model that was
time-reversible. The multiplicative fluctuation relations were studied 
recently in \cite{FouxHorv} also for particles with inertia carried by 
the homogeneous Kraichnan flow. Due to the Stokes friction force, 
the standard time-reversibility is broken in such a system, leading to 
a modification of the relation between the rate functions $\,Z'\,$ and $\,Z$. 
\vskip 0.3cm

\nsection{Towards $N$-point hierarchy of fluctuation relations}
\label{sec:Npt}

Another way to induce new diffusive processes from the original one
described by the SDE (\ref{E0}) is to consider simultaneously its $N$ 
solutions starting at different initial points. They may be viewed as
a solution of the SDE 
\qq
\dot{\bm{x}}\ =\ \bm{u}_t(\bm{x})\,+\bm{v}_t(\bm{x})
\label{EN}
\qqq
with $\,\bm{x}=(x_1,\dots,x_N)$, $\,\bm{u}_t(\bm{x})=(u_t(x_1),\dots,
u_t(x_N))$, \,and $\,\bm{v}_t(\bm{x})=(v_t(x_1),\dots,v_t(x_N))$. 
\,The covariance of the white noise vector field $\,\bm{v}_t\equiv(v_{t,1},
\dots,v_{t,N})\,$ appearing on the right hand side is
\qq
\langle\,v^i_{t,m}(\bm{x})\,v^j_{s,n}(\bm{y})\,\rangle\ =\ \delta(t-s)\,
\,D^{ij}_t(x_m,y_n)\,.
\nonumber
\qqq
The spatial part of the covariance restricted to the diagonal is
\qq
d^{ij}_{t,mn}(\bm{x})\ =\ D^{ij}_t(x_m,x_n)\,.
\nonumber
\qqq
The machinery producing the fluctuation relations described in this
paper may be applied to the $\,N$-point diffusion process governed by the SDE
(\ref{EN}), at least if the matrix $\,\big(d^{ij}_{t,mn}(\bm{x})\big)\,$ 
is invertible, recall that the inverse of the matrix $\,\big(d^{ij}_t(x)
\big)\,$ appears in the expression (\ref{js}) for $\,\CJ_t$.
\,We postpone a closer examination of the possible hierarchy 
of fluctuation relations obtained this way to the future. Here, let us 
only remark that the tangent process $\,(\sfx_t,\sfX_t)$, \,which was 
studied in the preceding section and led to the multiplicative fluctuation 
relation (\ref{MFR}), could be viewed as a limiting case of the 
$\,(d+1)$-point process where the last $\,d\,$ points are infinitesimally 
close to the first one.
\vskip 0.3cm

\nsection{Conclusions} 
\label{sec:concl}

We have developed a unified approach to fluctuation relations
for finite-dimensional diffusion processes. The setup of the paper
covered the cases of deterministic dissipative continuous-time 
dynamical systems, of the Langevin dynamics with non-conservative forces,
and of the Kraichnan model of hydrodynamic flows. The fluctuation 
relations were obtained by comparing the forward diffusion process
to the backward one produced by a time inversion. We have admitted
different time inversions that treated differently two parts of the 
deterministic drift in the diffusion equations. This was physically 
motivated in situations when one part of the drift was assimilated 
with a dissipative and another one with a conservative force, but 
was used in other situations as well, leading to a greater flexibility. 
As particular cases, we discussed the natural time inversion used 
for deterministic systems, its slight modification for stochastic 
dynamics that permitted to take easily the deterministic limit 
of fluctuation relations, as well as the reverse protocol and the 
current reversal discussed in a similar context in \cite{ChChJarz},
and the complete reversal. We showed that any of the allowed time 
inversions leads to a detailed fluctuation relation (\ref{DFR}) 
of Proposition 7 that may be viewed as a constrained version of the 
generalized detailed balance relation to which the relation
(\ref{DFR}) reduces in the case of the complete reversal. The constraint 
fixes the value of the entropy production measured relative to the 
corresponding backward process. We obtained various transient fluctuation 
relations as corollaries of the detailed one. Among examples were the 
Evans-Searles fluctuation relation (\ref{EvSear}), the Crooks one 
(\ref{Crooks}), and various versions of the Jarzynski equality 
(\ref{Jarz}), including the original ones for the deterministic Hamiltonian 
dynamics and for the Langevin dynamics with local detailed balance 
(\ref{Jarz1}), the one for reversed protocol, and the  Hatano-Sasa one. 
By comparing the detailed fluctuation relations for two different time 
inversions, we obtained also a generalization (\ref{SpSgen}) of the 
Speck-Seifert equality (\ref{SpSeif}). For the sake of completeness, 
we included into the paper a derivation from the Jarzynski equality 
of the Green-Kubo and the Onsager relations, and of the 
fluctuation-dissipation theorem. On a simple example of a one-dimensional 
Langevin equation with spontaneously broken equilibrium, we indicated how 
in such a situation the Hatano-Sasa version of the Jarzynski equality 
induced corrections to the fluctuation-dissipation theorem proportional 
to the flux of the probability current. 
\vskip 0.2cm

In the case of stationary diffusion processes, we pointed out that 
the transient fluctuation relations may give rise to the asymptotic 
symmetries of the large-deviations rate function of the entropy 
production which were established first by Gallavotti-Cohen for the 
uniformly hyperbolic dynamical systems and were extended later to 
(some) diffusion processes by Kurchan and Lebowitz-Spohn. Finally, 
we wrote explicitly a detailed fluctuation relation for the induced 
tangent diffusion process obtained from the original one. This 
produced a multiplicative transient fluctuation relation that led 
for long times to a Gallavotti-Cohen-type symmetry of the large-deviations 
rate function for the stretching exponents governing the behavior of 
infinitesimally close trajectories of the diffusion process. We 
speculated that considering distant multi-point trajectories of the 
process should give rise to a hierarchy of fluctuation relations. 
It could also provide a way to produce fluctuation relations for 
flow processes describing the simultaneous evolution of all trajectories 
of the process \cite{Kunita}. A similar extension should also permit 
to formulate fluctuation relations for hydrodynamic flows modeling 
fully developed turbulence \cite{GawWarw,LeJR}. We postpone such 
questions to further studies.

\appendix

\newcommand{\appsection}[1]{\let\oldthesection\thesection
  \renewcommand{\thesection}{Appendix \oldthesection}
  \section{#1}\let\thesection\oldthesection\setcounter{equation}{0}}

\appsection{}

\noindent The Stratonovich SDE (\ref{E0}) defining the process $\,\sfx_t\,$
is equivalent to the It\^o SDE
\qq
dx^i\,=\,\big(u^i_t(x)+\tilde u^i_t(x)\big)\,dt\,+\,v^i_t(x)\,dt\,,
\nonumber
\qqq
with the correction term
\qq
\tilde u^i_t(x)\,=\,\frac{_1}{^2}\partial_{x^j}D^{ij}_t(x,y)|_{y=x}\,.
\nonumber
\qqq
By the It\^o calculus, $\,g(\sfx_t)\,$ satisfies the It\^o SDE
\qq
dg(x)\,=\,\big(u^i_t(x)+\tilde u^i_t(x)\big)\partial_ig(x)\,dt\,+\,
v^i_t(x)\partial_ig(x)\,dt\,+\,\frac{_1}{^2}d^{ij}_t(x)
\partial_i\partial_jg(x)\,dt\,.
\nonumber
\qqq
with the second order It\^o term. For the expectation of $\,g(\sfx_t)$, 
\,this gives the ODE
\qq
\frac{d}{dt}\,\bE^{t_0}_x\,g(\sfx_t)\,=\,\bE^{t_0}_x\,\,\big(u^i_t(\sfx_t)+
\tilde u^i_t(\sfx_t)\big)\partial_ig(\sfx_t)\,+\,
\frac{_1}{^2}d^{ij}_t(\sfx_t)\partial_i\partial_jg(\sfx_t)
\nonumber
\qqq
from which the formula
\qq
L_t\,=\,\big(u^i_t+\tilde u^i_t(\sfx_t)\big)\partial_i\,+\,
\frac{_1}{^2}d^{ij}_t\partial_i\partial_j\,,
\nonumber
\qqq
easily seen to be equivalent to Eq.\,(\ref{Generator}), \,follows.
\vskip 0.7cm

\appsection{}

\noindent{\bf Proof of Lemma 1.}
\qq
(L_{t,-}Rg)(x)&=&u^i_{t,-}(x)\partial_i(Rg)(x)\,=\,u^i_{t,-}(x)\,
(\partial_i{x^*}^k)(x)(\partial_kg)(x^*)\cr&=&-(u'^k_{t^*}
\partial_kg)(x^*)\,=\,-(R\,L'_{t^*,-}g)(x)\,,\cr\cr
(L_{t,+}Rg)(x)&=&\hat u^i_{t,+}(x)\partial_i(Rg)(x)\,+\,\frac{_1}{^2}
\partial_jd^{ij}_t(x)\partial_i(Rg)(x)\cr
&=&u^i_{t,+}(x)(\partial_i{x^*}^k)(x)(\partial_kg)(x^*)\,-\,
\frac{_1}{^2}\partial_{y^j}D^{ij}(x,y)|_{y=x}\,(\partial_i{x^*}^k)(x)
(\partial_kg)(x^*)\cr
&&+\,\frac{_1}{^2}(\partial_j{x^*}^l)(x)
\partial_{{x^*}^l}d^{ij}_t(x)(\partial_i{x^*}^k)(x)(\partial_kg)(x^*)\cr
&=&(u'^k_{t^*,+}\partial_kg)(x^*)\,-\,
\frac{_1}{^2}(\partial_j{x^*}^l)(y)\partial_{{y^*}^l}D^{ij}(x,y)|_{y=x}
\,(\partial_i{x^*}^k)(x)(\partial_kg)(x^*)\cr
&&-\,\frac{_1}{^2}\big(\partial_{{x^*}^l}\partial_j{x^*}^l(x)\big)
\,d^{ij}_t(x)(\partial_i{x^*}^k)(x)(\partial_kg)(x^*)\cr
&&+\,\frac{_1}{^2}
\partial_{{x^*}^l}\,(\partial_j{x^*}^l)(x)d^{ij}_t(x)
(\partial_i{x^*}^k)(x)(\partial_kg)(x^*)\cr
&=&(u'^k_{t^*,+}\partial_kg)(x^*)\,-\,\frac{_1}{^2}\partial_{{y^*}^l}\,
(\partial_j{x^*}^l)(y)D^{ij}_t(x,y)(\partial_i{x^*}^k)(x)|_{y=x}\,
(\partial_kg)(x^*)\cr
&&+\,\frac{_1}{^2}\partial_{{x^*}^l}d'^{kl}_{t^*}(x^*)(\partial_kg)(x^*)\cr
&=&(\hat u'^k_{t^*,+}\partial_kg)(x^*)\,+\,\frac{_1}{^2}(\partial_{l}
d'^{kl}_{t^*}\partial_kg)(x^*)\,=\,(R\,L'_{t^*,+}g)(x)\,,
\nonumber
\qqq
where we have used the relations (\ref{stand}), (\ref{utou'}, (\ref{D'D})
and (\ref{stand'}).

\hspace{10cm}$\Box$

\vskip 0.6cm

\appsection{}

\noindent In order to prove the first of the equalities (\ref{hatback}),
let us note that the condition $\,\hat u_{t,+}=0\,$ means that
\qq
u^i_{t,+}(x)\,=\,\frac{_1}{^2}\partial_{y^j}D^{ij}_t(x,y)|_{y=x}
\nonumber
\qqq
so that, according to Eqs.\,(\ref{utou'}) and (\ref{D'D}),
\qq
{u'}^{i}_{t^*,+}(x^*)&=&(\partial_k{x^*}^{i})(x)\frac{_1}{^2}
\partial_{y^l}D^{kl}_{t^*}(x,y)|_{y=x}\cr
&=&\frac{_1}{^2}\partial_{y^n}\,(\partial_j{x^*}^n)(y^*)
(\partial_l{x^*}^j)(y)(\partial_k{x^*}^{i})(x)D^{kl}_t(x,y)|_{y=x}\cr
&=&\frac{_1}{^2}(\partial_j{x^*}^n)(y^*)\partial_{y^n}\,
(\partial_k{x^*}^{i})(x)D^{kl}_t(x,y)(\partial_l{x^*}^j)(y)|_{y=x}\cr
&&+\,\frac{_1}{^2}\big(\partial_{x^n}(\partial_j{x^*}^n)(x^*)\big)
(\partial_k{x^*}^{i})(x)d^{kl}_t(x)(\partial_l{x^*}^j)(x)\cr
&&=\frac{_1}{^2}\partial_{y^j}D'^{ij}_{t^*}(x^*,y)|_{y=x^*}\,+\,
\frac{_1}{^2}(\partial_n{x^*}^k)(x)(\partial_k\partial_j{x^*}^n)(x^*)\,
d'^{ij}_{t^*}(x^*)\cr
&&=\frac{_1}{^2}\partial_{y^j}D'^{ij}_{t^*}(x^*,y)|_{y=x^*}\,+\,
\frac{_1}{^2}d'^{ij}_{t^*}(x^*)(\partial_j\ln\sigma)(x^*)\,,
\nonumber
\qqq
where we used the identity
\qq
(\partial_j\ln\sigma)(x^*)=(\partial_n{x^*}^k)(x)
(\partial_j\partial_k{x^*}^n)(x^*)
\nonumber
\qqq
to obtain the last equality. The first of the relations in
Eqs.\,(\ref{hatback}) follows. The second one is an immediate
consequence of the transformation rule in Eqs.\,(\ref{utou'}). 
\vskip 0.7cm

\appsection{}

\noindent{\bf Proof of Lemma 2.} \ \ The Cameron-Martin-Girsanov 
formula\footnote{We have transformed the formula usually written in 
the It\^{o} convention \cite{StVar} to the Stratonovich one.} says that 
if $\,\sfy_t\,$ is the diffusion process solving the SDE
\qq
\dot{y}\ =\ w_t(y)\,+\,u_t(y)\,+\,v_t(y)\,,  
\label{EA}
\qqq
then 
\qq
\bE^{t_0}_x\,\CF(\sfy)\,=\,
\bE^{t_0}_x\,\CF(\sfx)\,\,\ee^{-U(\sfx)}\,
\nonumber
\qqq
for $\,\sfx_t\,$ solving the SDE (\ref{E0}) and
\qq
U(\sfx)\,=\,\int\limits_{t_0}^t\Big[-\,w_s
(\sfx_s)\cdot d_s^{-1}(\sfx_s)\,\dot{\sfx}_s\,+
\,w_s(\sfx_s)\cdot d_s^{-1}(\sfx_s)\,\hat u_s(\sfx_s)\cr
+\,\frac{_1}{^2}w_s(\sfx_s)\cdot d_s^{-1}(\sfx_s)\,
w_s(\sfx_s)\,+\,\frac{_1}{^2}(\nabla\cdot w_s)
(\sfx_s)\Big]ds
\nonumber
\qqq
if the functional $\,\CF\,$ depends on the process restricted to the
time interval $\,[t_0,t]$. \,The first term under the integral
in the expression for $\,U(\sfx)\,$ has to be interpreted with the 
Stratonovich rule. Denoting by $\,\tilde L_t\,$
the generator of the process $\,\sfy_t\,$ solving the SDE (\ref{EA}),
\qq
\tilde L_t\,=\,(w^i_t+\hat u^i_t)\partial_i+\frac{_1}{^2}
\partial_jd^{ij}_t\partial_i\,,
\nonumber
\qqq
we obtain this way the relation
\qq
\tilde P_{t_0,t}(x,dy)\,\equiv\,
\bE^{t_0}_x\,\,\delta(\sfy_t-y)\,dy
\,=\,\overrightarrow{\cal T}\,\exp\Big[
\int\limits_{t_0}^t\tilde L_s\,ds\Big](x,dy)
\,=\,\bE^{t_0}_x\,\ee^{-U(\sfx)}\,\delta(\sfx_t-y)\,dy\,.
\nonumber
\qqq
Next, if $\,f_t(x)\,$ is a time-dependent function then, by the 
Feynman-Kac formula,
\qq
\overrightarrow{\cal T}\,\exp\Big[
\int\limits_{t_0}^t(\tilde L_s-f_s)\,ds\Big](x,dy)
&=&\bE^{t_0}_x\,\ee^{-U(\sfx)-\int\limits_{t_0}^t
f_s(\sfx_s)\,ds}\,\delta(\sfx_t-y)\,dy\,.
\nonumber
\qqq
The application of the latter formula for $\,w_t=-2\hat u_{t,+}\,$
and $\,f_t=-\nabla\cdot\hat u_{t,+}+\nabla\cdot u_{t,-}\,$
gives Eq.\,(\ref{W0}) in view of the relation (\ref{L1}).
\vskip 0.2cm

\hspace{10cm}$\Box$
\vskip 0.7cm

\appsection{}

Here we show that the matrix $\,M\,$ given by Eq.\,(\ref{M}),
where $\,\Gamma\,$ and $\,C\,$ are strictly positive and $\,\Pi\,$
is antisymmetric, has eigenvalues with negative real parts
and that the matrix $\,C\,$ may be recovered from Eq.\,(\ref{Ct})
by setting $\,t=\infty$.
\,If $\,\lambda\,$ is an eigenvalue of $\,M$, \,i.e. if
\qq
-(\Gamma-\Pi)C^{-1}x_\lambda\,=\,\lambda x_\lambda
\nonumber
\qqq
for some $\,x_\lambda\not=0\,$ then
\qq
\lambda\,=\,\frac{-\,x_\lambda\cdot C^{-1}(\Gamma-\Pi)C^{-1}x_\lambda}
{x_\lambda\cdot C^{-1}x_\lambda}\,=\,
\frac{-\,x_\lambda\cdot C^{-1}\Gamma C^{-1}x_\lambda}
{x_\lambda\cdot C^{-1}x_\lambda}\,<\,0\,.
\nonumber
\qqq
Eq.\,(\ref{M}) implies that
\qq
MC\,+\,CM^T\,=\,-2\Gamma
\nonumber
\qqq
which is solved by $\,C_\infty\,$ given by Eq.\,(\ref{Ct}) 
with $\,t=\infty$. \,Besides, this is the unique 
solution because if $\,MD+DM^T=0\,$ then
\qq
\frac{d}{dt}\,\ee^{\,tM}D\ee^{\,tM^T}\,=\,0
\nonumber
\qqq
and 
\qq
D\ =\ \lim\limits_{t\to\infty}\,\,\ee^{\,tM}D\ee^{\,tM^T}\ =\ 0\,.
\nonumber
\qqq

\appsection{}

\noindent{\bf Proof of Proposition 2.}\ \ It is enough to check the 
last identity for the so called cylindrical functionals
\qq
\CF(\sfx)\,=\,f(\sfx_{t_1},\dots,\sfx_{t_n})
\nonumber
\qqq
for $\,0\leq t_1\leq\cdots\leq t_n\leq T$. \,Since
\qq
\ee^{-\CW}\,=\,\ee^{\varphi_0(\sfx_0)}\,
\ee^{-\int\limits_0^{t_1}\CJ_s\,ds}
\,\ee^{-\int\limits_{t_1}^{t_2}\CJ_s\,ds}\,\cdots\ \,
\ee^{-\int\limits_{t_n}^{T}\CJ_s\,ds}\,
\ee^{-\varphi_T(\sfx_T)}\,,
\nonumber
\qqq
then, in virtue of Eq.\,(\ref{W0}), the left hand side of 
Eq.\,(\ref{BI0}) is equal to
\qq
dx\,\int f(x_1\dots,x_n)\,P^1_{0,t_1}(x,dx_1)\,
P^1_{t_1,t_2}(x_1,dx_2)\,\,\cdots\,
\,P^1_{t_n,T}(x_n,dy)\,\,\ee^{-\varphi_T(y)}
\nonumber
\qqq
with the integral over $\,x_1,\cdots,x_n$. \,With the use of relation 
(\ref{P1P'k}), this may be rewritten as
\qq
\ee^{-\varphi_T(y)}dy\int f(x_1\dots,x_n)\,P'_{0,t_n^*}(y^*,dx_n^*)
\,\,\cdots\,\,P'_{t^*_2,t^*_1}(x_2^*,dx_1^*)\,
\,P'_{t^*_1,T}(x_1^*,dx^*)
\nonumber
\qqq
and, after the change of variables $\,x_{i+1}^*\mapsto x'_{n-i}$,
\,as
\qq
&&\ee^{-\varphi_T(y)}dy\,\int f(x'^*_n\dots,x'^*_1)
\,P'_{0,t_n^*}(y^*,dx'_1)\,\,
\cdots\,\,P'_{t^*_2,t^*_1}(x'_{n-1},dx'_n)\,\,P'_{t^*_1,T}(x'_n,dx^*)
\nonumber
\qqq
This is equal to the left hand side of the identity (\ref{BI0})
since $\,\ee^{-\varphi_T(y)}dy=\ee^{-\varphi'_0(y^*)}dy^*$. 

\hspace{10cm}$\Box$

\end{document}